\documentclass[draft]{article}
\def\today{20.06.05} 
\usepackage{amsmath,amsfonts,amsthm,amssymb,amscd}

\newcommand{\be}{\begin{equation}}
\newcommand{\ee}{\end{equation}}
\newcommand{\bs}{\begin{split}}
\newcommand{\es}{\end{split}}
\newcommand{\ba}{\begin{align}}
\newcommand{\ea}{\end{align}}

\theoremstyle{plain} \newtheorem{theorem}{Theorem}[section]
\newtheorem{lemma}[theorem]{Lemma}

\newtheorem{corollary}[theorem]{Corollary} 
\newtheorem{remark}[theorem]{Remark}

\newcommand{\R}{{\mathbb R}}

\newcommand{\E}{{\mathcal E}}

\newcommand{\A}{{\mathcal A}}
\newcommand{\B}{{\mathcal B}}

\newcommand{\F}{{\mathcal F}}
\newcommand{\M}{{\mathcal M}}
\newcommand{\C}{\mathbb{C}}

\newcommand{\esp}{E_{\sigma}(P)}
\newcommand{\psp}{\Phi_{\sigma}(P)}
\newcommand{\hp}{H_(P)}
\newcommand{\hsp}{H_{\sigma}(P)}
\newcommand{\hop}{H_{0}(P)}
\newcommand{\hisp}{H_{I\sigma}(P)}
\newcommand{\ths}{\widetilde H_{\sigma}}
\newcommand{\tho}{\widetilde H_{0}}
\newcommand{\this}{\widetilde H_{I\sigma}}
\newcommand{\emsp}{E_{\mathrm{mod},\sigma}(P)}
\newcommand{\hmsp}{H_{\mathrm{mod},\sigma}(P)}

\newcommand{\omod}{\omega_{\mathrm{mod}}(k)}

\newcommand{\om}{\Omega_{\rm ph}}
\newcommand{\rs}{\rho_{\sigma}}
\newcommand{\xt}{\tilde x}
\newcommand{\bel}[1]{\be\label{#1}}
\newcommand{\bea}{\begin{eqnarray}}
\newcommand{\beal}[1]{\begin{eqnarray}\label{#1}}
\newcommand{\eea}{\end{eqnarray}}
\newcommand{\bal}{\begin{align}}
\newcommand{\basl}[1]{\begin{align}\begin{split}\label{#1}}
\newcommand{\easl}{\end{split}{\end{align}}}
\newcommand{\lef}{\left}
\newcommand{\ej}{e^{-ik\cdot \tilde x_j}}
\newcommand{\el}{e^{-ik\cdot \tilde x_l}}
\newcommand{\en}{e^{-ik\cdot \tilde x_{N+1}}}
\newcommand{\emk}{\epsilon_\mu(k)}
\newcommand{\siun}{\sum_{i=1}^N}
\newcommand{\sjun}{\sum_{j=1}^N}
\newcommand{\skun}{\sum_{k=1}^N}
\newcommand{\slun}{\sum_{l=1}^N}
\newcommand{\ri}{\right}
\newcommand{\pmd}{\left( P-{\rm d}\Gamma(k)\right)}
\newcommand{\rt}{\tilde{R}}
\newcommand{\at}{\tilde{A}}
\newcommand{\usd}{\frac{1}{2}}
\newcommand{\rst}{\tilde{\rho}_{\sigma}}
\newcommand{\pstp}{\widetilde{\Phi}_{\sigma}(P)}
\newcommand{\pst}{\widetilde{\Phi}_{\sigma}}
\newcommand{\thsp}{\widetilde H_{\sigma}(P)}

\newcommand{\va}[1]{|#1|}
  \newcommand{\vak}{|k|}

  \newcommand{\vag}{|g|}
  
\newcommand{\norm}[1]{\left\Vert#1\right\Vert}

\numberwithin{equation}{section}

\begin{document}

\author{
Laurent AMOUR,%
\footnote{Laboratoire de Math\'ematiques, UMR-CNRS 6056,
Universit\'e de Reims,
 Moulin de la Housse - BP 1039,
 51687 REIMS Cedex 2, France. }
\quad  Beno\^\i t GR\'EBERT%
\footnote{Laboratoire de Math\'ematiques Jean LERAY, UMR-CNRS
6629, Universit\'e de Nantes, 2, rue de la Houssini\`ere, 44072
NANTES Cedex 03, France.}
\\ and \\ Jean-Claude GUILLOT%
\footnote{ Centre de Math\'ematiques Appliqu\'ees, UMR-CNRS 7641,
Ecole Polytechnique, 91128 Palaiseau Cedex, France.} }

\title{The dressed mobile atoms and ions}

\date{\today}
\maketitle

\begin{abstract}
We consider free atoms and ions in $\R^3$ interacting with the
quantized electromagnetic field. Because of the translation
invariance we consider the reduced hamiltonian associated with the
total momentum. After introducing an ultraviolet cutoff we prove
that the reduced hamiltonian for atoms has a ground state if the
coupling constant and the total momentum are sufficiently small.
In the case of ions an extra infrared regularization is needed. We
also consider the case of the hydrogen atom in a constant magnetic
field. Finally we determine the absolutely continuous spectrum of
the reduced hamiltonian.
\end{abstract}
\newpage
\tableofcontents
\newpage
\setcounter{section}{0}

\section{Introduction.}\label{s1}
In \cite{BFS99} V. Bach, J. Fr\"ohlich and M. Sigal consider atoms
and molecules with fixed nuclei interacting with the quantized
electromagnetic field. If the interaction between the electrons
and the quantized radiation field is turned off, the atom or
molecule is assumed to have at least one bound state. After
introducing an ultraviolet cutoff for the quantized field, they
prove that the interacting system has a ground state corresponding
to the bottom of the energy spectrum for sufficiently small values
of the fine structure constant. In \cite{GLL} and \cite{LL03}, M.
Griesemer, E. Lieb and M. Loss have been able to get rid of the
smallness condition concerning the fine structure constant (see also
\cite{BCV}).

 In this paper we consider free atoms and ions in
$\R^3$ interacting with the quantized electromagnetic field.
Because of the translation invariance in $\R^3$ we consider the
reduced hamiltonian associated with the total momentum for
particles and field. After introducing and ultraviolet cutoff for
the quantized electromagnetic field we prove that the reduced
hamiltonian for atoms has a ground state if the coupling constant
of the interaction between the particles and the field together
with the total momentum are sufficiently small. In the case of
ions we need to introduce an infrared regularization in order to
get the same result. The infrared regularization is not needed in
the case of atoms because we are able to use a simple form of the
Power-Zienau-Woolley transformation for a moving neutral system of
charges (see \cite{BFS99}, (\cite{GLL}, \cite{GR}). We also
consider a nonrelativistic hydrogen atom interacting with a
constant magnetic field and with a quantized electromagnetic
field. Again a Power-Zienau-Woolley transformation applies and we
prove that the reduced Hamiltonian associated with the total
momentum has a ground state under the same conditions as without a constant magnetic field.
 This result has to be compared with
\cite{AGG} where we consider an electron in a classical magnetic
field pointing along the $x_3$-axis interacting with a quantized
electromagnetic field. In this case we need an infrared
regularization because of the
 the system is charged.

 In the case of one free particle a similar problem has been
studied in \cite{Chen} where T. Chen considers a freely
propagating relativistic spinless charged particle interacting
with the quantized electromagnetic field.
 The one-particle sector of Nelson's model
has been studied first by J. Fr\"ohlich (see\cite{F73},
\cite{F74}) and more recently by A. Pizzo (see
\cite{piz03},\cite{piz04}) and J.S. M{\o}ller (see \cite{Mo}).

As in \cite{AGG} our proof combines the approach associated with
hamiltonian which are invariant by translation (see\cite{F73},
\cite{piz03}, \cite{piz04}, \cite{FGSray}) together with that
concerning confined system of charges (see\cite{BCV}, \cite{H01},
\cite{H99}, \cite{H00}, \cite{H04}, \cite{GLL}, \cite{LL03},
\cite{FGScom}).

Let us finally remark that the same result holds in the case of atoms moving in a waveguide and interacting with the quantized electromagnetic field in the wave guide (see \cite{AB}) and in the case of dressed atoms moving in a periodic potential (see \cite{Cohen}).
\section{Definition of the model and self-adjointness.}\label{s2}

\subsection{The Hamiltonian.}\label{s2.1}
We consider $N$ electrons in $\R^3$ of charge $e$ and mass $m$
interacting with a nucleus of charge $–Ze$, mass $m_{\rm ncl}$ and
spin $S$ and with photons. We suppose $N\geq 2$. The associated
Pauli Hamiltonian in Coulomb gauge is formally given by

\begin{align}
      \begin{split}\label{Hfor}
   H&\equiv  H_{N,Z}=\sum_{j=1}^N\frac{1}{2m}\left(p_{j}-eA(x_j)\right)^2
    +
    \frac{1}{2m_{\rm ncl}}\left(p_{N+1}+ZeA(x_{N+1})\right)^2\\
    &+V_{cl}(x)\otimes 1+
    1\otimes H_{\rm ph}-\frac{e}{2m}\sum_{j=1}^N\sigma_j \cdot B(x_j)
    +g_{\rm ncl}\frac{Ze}{2m_{\rm ncl}} S\cdot B(x_{N+1})
\end{split}\end{align}

with
\be
V_{cl}(x)\equiv V_{cl}(x_1,\ldots,x_{N+1}:=\sum_{j=1}^N\frac{Ze^2}{\vert x_j-x_{N+1}\vert}
    +
   \sum_{1\leq i<j\leq N}^N\frac{e^2}{\vert x_i-x_{j}\vert}\ .
\ee

Here the units are such that  $\hbar=c=1$, $p_j=-i\nabla_{x_j}$ ,
$x_j=({x_j}_1,{x_j}_2,{x_j}_3)$,
$\sigma_j=({\sigma_j}_1,{\sigma_j}_2,{\sigma_j}_3)$is the
3-component vector of the Pauli spin matrices for the $j$th
electron. $S=(S_1,S_2,S_3)$ is a $3-$component vector of spin
hermitian matrices in  $\C^d$  for the nucleus and $g_{\rm ncl}$
is the Land\'e factor of the nucleus.

The quantized electromagnetic field is formally given by

\be A(x)= \frac{1}{2\pi}\sum_{\mu =1,2} \int d^3k \left(
\frac{1}{\vak^{1/2}}\epsilon_{\mu}(k)e^{-ik\cdot
x}a^\star_{\mu}(k)\right. + \left.
\frac{1}{\vak^{1/2}}\epsilon_{\mu}(k)e^{ik\cdot x}a_{\mu}
(k)\right)\ , \ee
\begin{align}
      \begin{split}
B(x)= \frac{i}{2\pi}\sum_{\mu =1,2} \int d^3k& \left\{
-\vak^{1/2}\left(\frac{k}{\vak}\wedge\epsilon_{\mu}(k)\right)\right.
e^{-ik\cdot
x}a^\star_{\mu}(k) \\
&+  \left.
\vak^{1/2}\left(\frac{k}{\vak}\wedge\epsilon_{\mu}(k)\right)
e^{ik\cdot x}a_{\mu}(k)\right\}
\end{split}
  \end{align}
 where
$\epsilon_{\mu}(k)$ are real polarization vectors satisfying
$\epsilon_{\mu}(k)\cdot \epsilon_{\mu'}(k)=\delta_{\mu \mu'}$,
$k\cdot \epsilon_{\mu}(k)=0$; $a_{\mu}(k)$ and $a^\star_{\mu}(k)$
are the usual annihilation and creation operators acting in the
Fock space
$$
\mathcal F := \oplus_{n=0}^{\infty}
L^2(\R^{3},\C^2)^{\otimes^n_{s}}
$$
where $L^2(\R^{3},\C^2)^{\otimes^0_{s}}=\C$ and
$L^2(\R^{3},\C^2)^{\otimes^n_{s}}$ is the symmetric $n$-tensor
power of $L^2(\R^{3},\C^2)$ appropriate for Bose-Einstein
statistics. The annihilation and creation operators  obey the
canonical commutation relations ($a^\sharp=a^\star$ or $a$) \be
\label{ccr} [a^\sharp_{\mu}(k),a^\sharp_{\mu'}(k') ]=0 \quad
\mbox{et} \quad [a_{\mu}(k),a^\star_{\mu'}(k') ]=\delta_{\mu
\mu'}\delta(k-k')\ . \ee Finally the Hamiltonian for the photons
in the Coulomb gauge is given by \be \label{Hphoton} H_{\rm ph}=
\sum_{\mu=1,2}\int \vak a_{\mu}^\star(k) a_{\mu}(k) d^3k\ . \ee
The Hilbert space associated with $H_{N,Z}$ is then
$$
\mathcal H={\cal A}_N\left[L^2(\R^3,\C^2)^{\otimes^N}\right]
\otimes L^2(\R^3,\C^d)\otimes\mathcal F.
$$
Here ${\cal A}_N$ is the orthogonal projection onto the subspace
of totally antisymmetric wave functions, as required by the Pauli
principle.

As it stands, the Hamiltonian $H$ cannot be defined as a
self-adjoint operator in $\mathcal H$ and we need to introduce
cutoff functions, both in $A(x)$ and in $B(x)$, which will satisfy
appropriate hypothesis in order to get a self-adjoint operator in
$H$.

This operator, still denoted by $H$, commutes with each
component of the total momentum of the system       denoted by
$P$. We have $P=\left(\sum_{j=1}^{N+1}p_j\right)\otimes 1+
1\otimes {\rm d}\Gamma(k)$ where, for every $i$, ${\rm
d}\Gamma(k_i)$ is the second quantized operator associated to the
multiplication operator by $k_i$ in $L^2(\R^3,\C^2)$. The joint
spectrum of $(P_1,P_2,P_3)$ is the real line. It turns out that
$H$ admits a decomposition over the joint spectrum of
$(P_1,P_2,P_3)$ as a direct integral :
\[
H\ \simeq\ \int_{\R^3}^\oplus H(P)\,\,d^3 P
\]
on
\[
\mathcal H\ \simeq\ \int_{\R^3}^\oplus \mathcal H(P)\,\,d^3P
\]
where \[ \mathcal H(P)=\mathcal A_N \left[
L^2(\R^{3},{\C^2})^{\otimes^N} \right]\otimes \C^d\otimes\F\] for
every $P\in\R^3$. The reduced operator $H(P)$ will be
explicitly computed and the aim of this article is to initiate the
spectral analysis of $H(P)$  when $\va{P}$ is small. We now
introduce the hamiltonian in $\mathcal H$ associated to
$(\ref{Hfor})$. As usual we will consider the charge $e$ in front
of the quantized electromagnetic fields $A(x)$ and $B(x)$ as a
parameter and from now on we will denote it by $g$. We introduce
$\rho(k)$ a cutoff function associated with an ultraviolet cutoff.
We suppose that $\rho(k)$ satisfies \be\label{6} \int_{\vak\leq
1}\frac{\va{\rho(k)}^2}{\vak^2} d^3 k+\int_{\vak\geq
1}\vak\va{\rho(k)}^2 d^3 k<\infty.\ee The associated quantized
electromagnetic field is then given by ($j=1,2,3$)
\begin{eqnarray} \label{Aj}
A_{j}(x, \rho)= \frac{1}{2\pi}\sum_{\mu =1,2} \int d^3k \left(
\frac{\rho(k)}{\vak^{1/2}}\epsilon_{\mu}(k)_{j}e^{-ik\cdot
x}a^\star_{\mu }(k)\right. \\ \nonumber + \left.
\frac{\bar{\rho}(k)}{\vak^{1/2}}\epsilon_{\mu}(k)_{j}e^{ik\cdot
x}a_{\mu} (k)\right)\ ,
\end{eqnarray}
\begin{eqnarray} \label{Bj}
B_{j}(x, \rho)= \frac{i}{2\pi}\sum_{\mu =1,2} \int d^3k \left(
-\vak^{1/2}\rho(k)\left(\frac{k}{\vak}\wedge\epsilon_{\mu}(k)\right)_{j}
e^{-ik\cdot x}a^\star_{\mu} (k)\right.\\ \nonumber + \left.
\vak^{1/2}\bar{\rho}(k)\left(\frac{k}{\vak}\wedge
\epsilon_{\mu}(k)\right)_{j} e^{ik\cdot x}a_{\mu} (k)\right)\ .
\end{eqnarray}

$H$ is then the following operator: \basl{9}
    H&=\sum_{j=1}^N\frac{1}{2m}\left(p_{j}-gA(x_j,\rho)\right)^2
    +
    \frac{1}{2m_{\rm ncl}}\left(p_{N+1}+ZgA(x_{N+1},\rho)\right)^2\\
    &+V_{cl}(x)\otimes 1
    1\otimes H_{\rm ph}-\frac{g}{2m}\sum_{j=1}^N\sigma_j\cdot B(x_j)
    +g_{\rm ncl}\frac{Zg}{2m_{\rm ncl}} S\cdot B(x_{N+1})
\end{split}\end{align}

Let $\F_{0,fin}$  be the set of  $(\psi_n)_{n\geq 0}\in\F$  such
that $\psi_n$ is in the Schwartz space for every $n$ and
$\psi_n=0$ for all but finitely many $n$. Then our model is
described by the operator $H$ as defined on
\[\mathcal H_{0,fin}=
 \mathcal A_N \left[
C^\infty_0(\R^{3},{\C^2})^{\otimes^N}\right] \otimes
C^\infty_0(\R^{3},{\C^d})\otimes \F_{0,fin}.
\]

The operator $H$ is then symmetric. In order to compute the
reduced hamiltonian $H(P)$ for a given value of the total
momentum $P$ we have to introduce the momentum of the center of
mass of the electrons the nucleus and, consequently, the Jacobi
variables for the atom.

We set \beal{6bis}
r_i&=&x_{i+1}-\frac{1}{i}(x_1+x_2+\cdots+x_i),\quad i=1,2,\dots,N
\\
\label{7} r_{N+1}&=&\frac{1}{M}\Bigl(m(x_1+x_2+\cdots+x_N)+m_{\rm
ncl}x_{N+1}\Bigr), \eea where \be\label{8}M=Nm+m_{\rm ncl}\ee is
the total mass.

Accordingly the corresponding canonically conjugate momentum
operators are defined by
\be\label{9bis}\omega_j=\frac{1}{i}\frac{\partial}{\partial
r_j},\quad j=1,2,\cdots,N+1\ee

with \be\label{10}\omega_{N+1}=\sum_{j=1}^{N+1}p_j.\ee

Set \be\label{11} X=(x_1,x_2,\dots,x_{N+1})\ee
\be\label{12}R=(r_1,r_2,\dots,r_{N+1}).\ee We have \be\label{13}
X=\A R\ee where $\A$ is a  $(N+1)\times(N+1)$  invertible matrix.

Therefore we have \be\label{14} x_j=(\A R)_j,\quad
j=1,2,\dots,N+1.\ee

Similarly we set
\be\label{16}P_X=(p_1,p_2,\dots,p_{N+1})\ee
\be\label{17}P_R=(\omega_1,\omega_2,\dots,\omega_{N+1})\ee

and we have \be P_X=\B P_R\label{17b}\ee where $\B$ is a
$(N+1)\times(N+1)$ invertible matrix.Notice that in fact,  $\B^{-1}={}^t\A$.

We thus get \be\label{18}p_j=(\B P_R)_j,\quad j=1,2,\dots,N+1\ee

with

\[ \omega_{N+1}=\sum_{l=1}^{N+1} p_l=\left(\B^{-1}
P_X\right)_{N+1}.\] Therefore

\be\label{19}\A_{j,N+1}=\B^{-1}_{N+1,j}=1,\quad j=1,2,\dots,N+1.\ee Let
$\M(x,\rho)$ be the following operator-valued vector

\be\label{20}\M(x\rho)=\left(gA(x_1,\rho),\dots,gA(x_N,\rho),-Zg
A(x_{N+1},\rho)\right)\ee

 i.e.,

\bea\nonumber  \M(x,\rho)_j&=&g A(x_j,\rho),\quad j=1,2,\dots,N\\
\M(x,\rho)_{N+1}&=&-Zg A(x_{N+1},\rho).\eea

 Let $\mu_i>0$ be the
reduced mass defined by

\be\label{21}\mu_i=\left(\frac{i}{i+1}\right)m,\quad
i=1,2,\dots,N.\ee We have

 \basl{22}
\sum_{j=1}^N\frac{1}{2m}&\left(p_{j}-gA(x_j,\rho)\right)^2
    +
    \frac{1}{2m_{\rm ncl}}\left(p_{N+1}+Zg A(x_{N+1},\rho)\right)^2\\
&=\sum_{j=1}^N\frac{1}{2m}\left((\B P_R)_{j}-gA((\A
R)_j,\rho)\right)^2\\
    &+
    \frac{1}{2m_{\rm ncl}}\left((\B P_R)_{N+1}+Zg A((\A R)_{N+1},\rho)\right)^2\\
&=\sum_{j=1}^N\frac{1}{2m}\left(\left[\B\left( P _R-{\B}^{-1}
\M({\A} R,\rho)\right)\right]_j\right)^2\\
    &+
    \frac{1}{2m_{\rm ncl}}\left(\left[\B\left( P _R-{\B}^{-1}
\M({\A} R,\rho)\right)\right]_{N+1}\right)^2.
\end{split}\end{align}
Now because of the Jacobi coordinates and because, in the
« Schr\"odinger representation » of Fock space $\mathcal F$,
$A(x,\rho)$ can be treated as a classical vector potential, we get

\basl{23}
\sum_{j=1}^N\frac{1}{2m}&\left(p_{j}-gA(x_j,\rho)\right)^2
    +
    \frac{1}{2m_{\rm ncl}}\left(p_{N+1}+Zg A(x_{N+1},\rho)\right)^2\\
&=\sum_{j=1}^N\frac{1}{2\mu_j}\left(\left( P _R-{\B}^{-1}
\M({\A} R,\rho)\right)_j\right)^2\\
    &+
    \frac{1}{2M}\left(\left( P _R-{\B}^{-1}
\M({\A} R,\rho)\right)_{N+1}\right)^2.
\end{split}\end{align}

Recall \[V_{\rm cl}(x)=
 -\sum_{j=1}^N\frac{Ze^2}{\vert x_j-x_{N+1}\vert}
    +
    \sum_{1\leq i<j\leq N}\frac{e^2}{\vert x_i-x_{j}\vert}\]

and set \be\label{24}\widetilde{V}_{\rm cl}(R)=V_{\rm cl}(\A R).\ee
Note that $\widetilde{V}_{\rm cl}(R)$ does not depend on
$r_{N+1}$.

By $(\ref{23})$ $H$ is unitarily equivalent to the following
operator defined on $
 \mathcal H_{0,fin}$
 and still denoted by $H$ :
\basl{25} H &=
     \frac{1}{2M}\left(\omega_{N+1}-\left( {\B}^{-1}
\M({\A} R,\rho)\right)_{N+1}\right)^2\\
& + \sum_{j=1}^N\frac{1}{2\mu_j}\left( \omega_j-\left({\B}^{-1}
\M({\A} R,\rho)\right)_j\right)^2\\
&+\widetilde{V}_{\rm cl}(R)\otimes 1 + 1\otimes H_{\rm ph}\\
&-\frac{g}{2m}\sum_{j=1}^N\sigma_j \cdot B((\A R)_j)
    +g_{\rm ncl}\frac{Zg}{2m_{\rm ncl}} S\cdot B((\A R)_{N+1})\\&
\end{split}\end{align}

We now want to show that $H_{N,Z}$ is essentially self adjoint on
$\mathcal H_{0,fin}$ when $g$ and $\rho(k)$ satisfy appropriate
conditions.

Set \be\label{27} b_{ij}=(\B^{-1})_{ij}.\ee

According to $(\ref{19})$, $b_{N+1,j}=1$, $j=1,2,\dots, N+1$. We
have \be\label{28}H=H_{0}+H_{I}\ee

with
\be\label{29}H_{0}=\left(\frac{\omega_{N+1}^2}{2M}+\sum_{j=1}^N
\frac{\omega_{j}^2}{2\mu_j}+\widetilde{V}_{\rm
cl}(R)\right)\otimes 1 +1\otimes H_{\rm ph}\ee

and \basl{30} H_{I} &=-\frac{g}{M}\sum_{j=1}^N A((\A
R)_{i},\rho)\cdot\omega_{N+1} +
    \frac{Zg}{M}
A(({\A} R)_{N+1},\rho)\cdot\omega_{N+1}\\
 &-g
\sum_{j=1}^N\frac{1}{\mu_j}\left( \sum_{k=1}^N b_{jk}A((\A
R)_{k},\rho)\cdot\omega_{j}\right)\\
&+Zg \sum_{j=1}^N \frac{1}{\mu_j}b_{j,N+1}A((\A
R)_{N+1},\rho)\cdot\omega_{j} \\
&-\frac{g}{2m}\sum_{j=1}^N \sigma_j\cdot B((\A R)_{j},\rho) +
   Zg \frac{g_{\rm ncl}}{2m_{\rm ncl}}
S\cdot B(({\A} R)_{N+1},\rho) \\
&+\frac{1}{2M}\left(\sum_{i=1}^N g A((\A R)_{i},\rho)-Zg A(({\A}
R)_{N+1},\rho)\right)^2\\
&+ \sum_{j=1}^N\frac{1}{2\mu_j}\left( \sum_{k=1}^N b_{jk} g A((\A
R)_{k},\rho)- b_{j,N+1} Zg A((\A R)_{N+1},\rho)\right)^2
\end{split}\end{align}

where we used $\varepsilon_\mu(k)\cdot k=0$.

One checks that \be\label{31} \sup_{i,j:i\not =N+1}
\va{b_{ij}}=1.\ee

One easily shows that, for $\psi\in \mathcal
A_N\left[C^\infty_0(\R^{3},{\C^2})^{\otimes^N}\right] \otimes
C^\infty_0(\R^{3},{\C^d})\otimes \F_{0,fin}$, \basl{32}
&\frac{\va{g}}{M}\norm{\left(\sum_{i=1}^N A((\A
R)_{i},\rho)\cdot\omega_{N+1}\right)\psi}\\
&\leq\frac{12\vag N}{\pi\sqrt{2M}}\left(
\int_{\R^3}\frac{\va{\rho(k)}^2}{\vak^2} d^3
k\right)^\frac{1}{2}\norm{\left(H_{0}-E_{elec}\right)\psi}+o(1)
\end{split}\end{align}
 where $o(1)$   are terms associated with operators
which are relatively bounded with respect to $H_{0,N,Z}$ with a
zero relative bound.

Similarly, \basl{33} &\frac{Z\va{g}}{M}\norm{\left( A((\A
R)_{N+1},\rho)\cdot\omega_{N+1}\right)\psi}\\
&\leq\frac{12\vag Z}{\pi\sqrt{2M}}\left(
\int_{\R^3}\frac{\va{\rho(k)}^2}{\vak^2} d^3
k\right)^\frac{1}{2}\norm{\left(H_{0}-E_{\rm
elec}\right)\psi}+o(1).
\end{split}\end{align}

Here  $E_{\rm elec}$  is the infimum of the spectrum of the
confining electronic part of $H_{0,N,Z}$, i.e.,   of
\[
\sum_{j=1}^N \frac{\omega_{j}^2}{2\mu_j}+\widetilde{V}_{\rm
cl}(R).
\]

In th same way we verify \basl{34} &\vag\norm{\left(
\sum_{j=1}^N\frac{1}{\mu_j}\left( \sum_{k=1}^N
b_{jk}A((\A R)_{k},\rho)\cdot\omega_{j}\right)\right)\psi}\\
&\leq\frac{12\vag  N^2}{\pi\sqrt{2\mu}}\left(
\int_{\R^3}\frac{\va{\rho(k)}^2}{\vak^2} d^3
k\right)^\frac{1}{2}\norm{\left(H_{0}-E_{\rm
elec}\right)\psi}+o(1),
\end{split}\end{align}

and

\basl{35} &Z \vag\norm{\left( \sum_{j=1}^N
\frac{1}{\mu_j}b_{j,N+1}A((\A
R)_{N+1},\rho)\cdot\omega_{j}\right)\psi}\\
&\leq\frac{12\vag Z  N}{\pi\sqrt{m}}\left(
\int_{\R^3}\frac{\va{\rho(k)}^2}{\vak^2} d^3
k\right)^\frac{1}{2}\norm{\left(H_{0}-E_{\rm
elec}\right)\psi}+o(1),
\end{split}\end{align}

where we used  \[\inf_j\mu_j=\frac{m}{2}.\]

In $(\ref{32})$, $(\ref{33})$, $(\ref{34})$ and $(\ref{35})$ we
have used the following well known estimates:

\be\nonumber\norm{a_{\mu}(g(.,x))\psi}\leq \left(\int
\frac{\va{g(x,k)}^2}{ \vak}d^3k\right)^{1/2}\norm{(I\otimes H_{\rm
ph}^{1/2})\psi} \ee and
\begin{align}\begin{split}\label{36}
\norm{a_{\mu}^*(g(.,x))\psi}&\leq \left(\int \frac{\va{g(x,k)}^2}{
\vak}d^3k\right)^{1/2}\norm{(I\otimes H_{\rm ph}^{1/2})\psi}\\
&+\left(\int {\va{g(x,k)}^2}d^3k\right)^{1/2}\norm{\psi}\ .
\end{split}\end{align}
Note that \basl{37}&\frac{\vag}{2m}\norm{\left(\sum_{j=1}^N
\sigma_j\cdot
B((\A R)_{j},\rho) \right)\psi}=o(1)\\
   &Z\vag \frac{g_{\rm ncl}}{2m_{\rm ncl}}
\norm{\left(S\cdot B(({\A} R)_{N+1},\rho) \right)\psi}=o(1).
\end{split}\end{align}
It remains to estimate the quadratic terms. Let us recall the
following estimates (c.f. \cite{A90})
\begin{align} \begin{split}\label{38}
\norm{a_{\mu}(f)a_{\lambda}(f)\psi}&\leq \left( \int
\frac{\va{\rho(k)}^2}{\vak^{2}}d^3k\right) \norm{(H_{\rm ph}
+1)\psi}\\  &+K\left( \int
\frac{\va{\rho(k)}^2}{\vak^{2}}d^3k\right)^{1/2} \left(\int
{\va{\rho(k)}^2}d^3k\right)^{1/2}
\norm{(H_{\rm ph} +1)^{1/2}\psi}\ ,\\
\norm{a^*_{\mu}(f)a_{\lambda}(f)\psi}&\leq \left( \int
\frac{\va{\rho(k)}^2}{\vak^{2}}d^3k\right)
\norm{(H_{\rm ph} +1)\psi}\\
&+\left( K\left( \int
\frac{\va{\rho(k)}^2}{\vak^{2}}d^3k\right)^{1/2}
\left( \int {\va{\rho(k)}^2}d^3k\right)^{1/2}\right.\\
&\left. +\left( \frac{\va{\rho(k)}^2}{\vak^{2}}d^3k\right)^{1/2}
\left(\frac{\va{\rho(k)}^2}{\vak}d^3k\right)^{1/2}\right)
\norm{(H_{\rm ph} +1)^{1/2}\psi}\ ,
\\\norm{a^*_{\mu}(f)a^*_{\lambda}(f)\psi}&\leq
\left( \int \frac{\va{\rho(k)}^2}{\vak^{2}}d^3k\right)
\norm{(H_{\rm ph} +1)\psi}\\
&+\left( K\left( \int
\frac{\va{\rho(k)}^2}{\vak^{2}}d^3k\right)^{1/2}
\left( \int {\va{\rho(k)}^2}d^3k\right)^{1/2}\right.\\
&\left. +\left( \frac{\va{\rho(k)}^2}{\vak^{2}}d^3k\right)^{1/2}
\left(\frac{\va{\rho(k)}^2}{\vak}d^3k\right)^{1/2}\right)
\norm{(H_{\rm ph} +1)^{1/2}\psi}\\
&\!\!\!\!\!\!\!\!\!\!\!\!\!\!+\left( \left( \int
\frac{\va{\rho(k)}^2}{\vak^{2}}d^3k\right)^{1/2} \left(\int \vak
{\va{\rho(k)}^2}d^3k\right)^{1/2} +\int
\frac{\va{\rho(k)}^2}{\vak}d^3k\right)\norm{\psi}
\end{split}\end{align}
where $K=\frac{1}{\pi}\int_{0}^\infty \frac{\sqrt
\lambda}{(1+\lambda)^2}d\lambda$.

As $(H_{\rm ph}+1)^\frac{1}{2}$ is relatively bounded with respect
to $H_{\rm ph}+1$ (and thus to $(H_{0}-E_{\rm elec})$) we
deduce that the relative bound of the quadratic terms in
$A(\cdot,\rho)$ of $(\ref{30})$ with respect to $H_{0}-E_{\rm
elec}$ is estimated by \be\label{39}
\left(\frac{2g^2}{M\pi}(N+Z)^2+\frac{4g^2}{m\pi}(N^3+2ZN^2+Z^2)\right)\int
\frac{\va{\rho(k)}^2}{\vak}d^3k.\ee

 Note that
$H_{0}$ is essentially self adjoint on  $\mathcal A_N \left[
C^\infty_0(\R^{3},{\C^2})^{\otimes^N}\right] \otimes\newline
C^\infty_0(\R^{3},{\C^d})\otimes \F_{0,fin}$      . Therefore, by
$(\ref{32})$, $(\ref{33})$, $(\ref{34})$, $(\ref{35})$,
$(\ref{37})$ and $(\ref{39})$, we get the following theorem from
the Kato-Rellich theorem.

\begin{theorem}\ \newline\label{t1}
 Assume \eqref{6} and
\basl{40}
&\frac{6\vag}{\pi}(N+Z)\left(\frac{1}{\sqrt{2M}}+\frac{1}{\sqrt{m}})\right)\left(\int_{\R^3}
\frac{\va{\rho(k)}^2}{\vak}d^3k\right)^\frac{1}{2}\\
&+
\frac{g^2}{\pi}\left(\frac{(N+Z)^2}{M}+\frac{2(N^2+Z)^2}{{m}})\right)\left(\int_{\R^3}
\frac{\va{\rho(k)}^2}{\vak}d^3k\right)<\frac{1}{2}.\end{split}\end{align}
Then $H$ is a self-adjoint operator in $\mathcal H$  with
domain $D(H)=D(H_{0})$ and $H$ is essentially
self-adjoint on $\mathcal A_N \left[
C^\infty_0(\R^{3},{\C^2})^{\otimes^N}\right] \otimes
C^\infty_0(\R^{3},{\C^d})\otimes$ $\F_{0,fin}$.\end{theorem}

\subsection{The reduced Hamiltonian.}\label{s2.2}
The operator $H$ is invariant by translation. Thus denoting
by $P$ the total momentum, i.e., $P=\omega_{N+1}\otimes 1+1\otimes
{\rm d}\Gamma(k)$, $H$ admit a decomposition over the joint
spectrum of $(P_1,P_2,P_3)$ as a direct integral
\[
H\simeq\int_{\R^3}^\oplus H(P)\,d^3 P
\]
on
\[
\mathcal H\simeq\int_{\R^3}^\oplus \mathcal H(P) \, d^3P
\]
where
\[
\mathcal H(P)=\mathcal A_N \left[ L^2(\R^{3},{\C^2})^{\otimes^N}
\right]\otimes \C^d\otimes\F.
\]
 To
compute $H(P)$ we proceed as in \cite{FGScom} and
\cite{A00}. Let $\Pi$ be the unitary map from$\mathcal H$ to
$L^2\left(\R^3,\mathcal A_N \left[ L^2(\R^{3},{\C^2})^{\otimes^N}
\right]\otimes \C^d\otimes\F\right)$ defined by \basl{41}
\left(\Pi\psi\right)_n\left(r_1,r_2,\dots,r_N,P,(k_1,\mu_1),(k_2,\mu_2),\dots,(k_n,\mu_n)\right)\\
=\widehat{\psi}_n\left(r_1,r_2,\dots,r_N,P-\sum_{i=1}^n
k_i,(k_1,\mu_1),(k_2,\mu_2),\dots,(k_n,\mu_n)\right)
\end{split}\end{align}

where  $\widehat{\psi}_n$ is the partial Fourier transformation
with respect to $r_{N+1}$. One easily verifies that, on $\mathcal
A_N \left[ C^\infty_0(\R^{3},{\C^2})^{\otimes^N}\right] \otimes
C^\infty_0(\R^{3},{\C^d})\otimes \F_{0,fin}$, \basl{42}
\Pi \omega_{N+1}\Pi^\star=P-{\rm d}\Gamma(k)\\
\Pi A_j\left((\A R)_l,\rho\right)\Pi^\star=A_j\left((\A \tilde{R})_l,\rho\right)\\
\Pi B_j\left((\A R)_l,\rho\right)\Pi^\star=B_j\left((\A
\tilde{R})_l,\rho\right)
\end{split}\end{align}

where \be\label{43}\widetilde{R}=(r_1,r_2,\dots,r_N,0).\ee
We remark that $(\A \tilde{R})_k= x_k -r_{N+1}$ for $j=1,\ldots , N+1$ 
represents the relative postion of the particle $k$ with respect to the center of mass.

 Then,
for $\psi\in\mathcal C^\infty_0(\R^{3})\otimes \mathcal A_N \left[
C^\infty_0(\R^{3N},{\C^2}^{\otimes^N})\right]\otimes\C^d \otimes
\F_{0,fin}$ we have \be\label{44}\left(\Pi
H\Pi^\star\psi\right)(P,\cdot)=H(P)\psi(P,\cdot)\ee

where the reduced hamiltonian $H(P)$ is given by
\be \label{45} H(P)=H_0(P)+H_I(P)\ee with
\basl{46} H_{0}&=\sum_{j=1}^N
\frac{\omega_{j}}{2\mu_j}+\widetilde{V}_{\rm
cl}(R) \otimes 1 +\\
&1\otimes\left\{\frac{1}{2M}\pmd^2+H_{\rm ph}\right\}
\end{split}\end{align}
and

\basl{47} H_I(P)=&-\frac{g}{M}\sum_{i=1}^N A((\A
\rt)_{i},\rho)\cdot\pmd \\
&+
    \frac{Zg}{M}
A(({\A} \rt)_{N+1},\rho)\cdot\pmd\\
 &-g
\sum_{j=1}^N\frac{1}{\mu_j}\left( \sum_{k=1}^N b_{jk}A((\A
\rt)_{k},\rho)\cdot\omega_{j}\right)\\
&+Zg \sum_{j=1}^N b_{j,N+1}A((\A
\rt)_{N+1},\rho)\cdot\omega_{j} \\
&-\frac{g}{2m}\sum_{j=1}^N \sigma_j \cdot B((\A \rt)_{j},\rho) +
   Zg \frac{g_{\rm ncl}}{2m_{\rm ncl}}
S\cdot B(({\A} \rt)_{N+1},\rho) \\
&+\frac{1}{2M}\left(\sum_{i=1}^N g A((\A \rt)_{i},\rho)-Zg A(({\A}
\rt)_{N+1},\rho)\right)^2\\
&+ \sum_{j=1}^N\frac{1}{2\mu_j}\left( \sum_{k=1}^N b_{jk} g A((\A
\rt)_{k},\rho)- b_{j,N+1} Zg A((\A \rt)_{N+1},\rho)\right)^2.
\end{split}\end{align}

 For every $P\in\R^3$, $H(P)$ is an operator in $\mathcal A_N \left[
L^2(\R^{3N},{\C^2}^{\otimes^N} \right]\otimes \C^d\otimes\F$ . We
want to show that this operator has a self-adjoint extension under
appropriate conditions on $\rho(\cdot)$ and $g$.

 The operator
$\left(\frac{1}{2M}\pmd^2+H_{\rm ph}\right)$ is essentially
self-adjoint on $\F_{0,fin}$. Therefore, for every $P\in\R^3$,
$H_{0}(P)$ is essentially self-adjoint on  \newline$\mathcal
A_N \left[ C_0^\infty(\R^{3},{\C^2})^{\otimes^N} \right]\otimes
\C^d\otimes\F_{0,fin}$. $H_{0}(P)$ still denotes its
self-adjoint extension. On the other hand, $H_{I}(P)$ is a
symmetric operator on \newline$\mathcal A_N \left[
C_0^\infty(\R^{3},{\C^2})^{\otimes^N} \right]\otimes
\C^d\otimes\F_{0,fin}$ and we want to prove that it is relatively
bounded with respect to $H_{0}(P)$. For that, by $(\ref{34}),
(\ref{35}), (\ref{37}), (\ref{39})$, we only need to estimates the
two first terms of the right hand side of $(\ref{47})$. For
$\psi\in \mathcal A_N \left[ C_0^\infty(\R^{3},{\C^2})^{\otimes^N}
\right]\otimes \C^d\otimes\F_{0,fin}$, one easily shows that
\basl{48} &\frac{\va{g}}{M}\sum_{i=1}^N\norm{\left( A((\A
\rt)_{i},\rho)\cdot\pmd\right)\psi}\\
&+ \frac{Z\va{g}}{M}\norm{\left( A((\A
\rt)_{N+1},\rho)\cdot\pmd\right)\psi}\\
&\leq\frac{12\vag
(N+Z)}{\pi}\left(\frac{1}{\sqrt{2M}}+\frac{1}{\sqrt{2\mu}}\right)
\norm{\left(H_{0}(P)-E_{\rm elec}\right)\psi}+o(1).
\end{split}\end{align}

Therefore we get
\begin{theorem}\ \newline\label{t2}  Assume $(\ref{6})$ and
$(\ref{40})$. Then, for every $P\in\R^3$, $H(P)$ is a
self-adjoint operator in $\mathcal
A_N\left[L^2(\R^{3},{\C^2})^{\otimes^N} \right]\otimes
\C^d\otimes\F$ with domain $D(H(P))=D(H_{0}(P))$ and
$H(P)$ is essentially self-adjoint on $\mathcal A_N\left[
C_0^\infty(\R^{3},{\C^2})^{\otimes^N} \right]$ $\otimes
\C^d\otimes\F_{0,fin}$.
\end{theorem}

Furthermore we get \begin{corollary}\ \newline\label{c3} We have
\[
\Pi H\Pi^\star=\int_{\R^3}^\oplus H(P)\,d^3 P
\]\end{corollary}

The proof of Corollary \ref{c3} follows by mimicking \cite{A00}.
\section{Main results.}\label{s3}

 Our main result states that, for $\va{P}$ and $\vag$
sufficiently small, $H(P)$ has a ground state.

But we have to distinguish the case of atoms, i.e., the case where $Z=N$
from the case of positive ions, i.e., the case
where $N<Z$. The problem is connected with the necessity  or not of
an infrared regularization of the cutoff function in order to
prove the existence of a ground state for $H$. In \cite{AGG}
such an infrared regularization of the cutoff function has been
introduced. But, in the case of atoms, we are able to use the
Power-Zienau-Woolley transformation (see \cite{GR}) in order to
get rid of any infrared regularization.

Let $h$ be the following self-adjoint operator in $\mathcal A_N
\left[L^2(\R^{3N},{\C^2}^{\otimes^N})\right]\otimes\C^d$
\begin{eqnarray}\label{49}
h=\sum_{j=1}^N \frac{\omega_{j}^2}{2\mu_j}+\widetilde{V}_{\rm
cl}(\tilde R).
\end{eqnarray}
In order to prove theorems \ref{t4}and \ref{t5} below  we only use the fact that 
$\inf\sigma(h)$is an
isolated eigenvalue of finite multiplicity. Recalling that ${V}_{\rm
cl}(R)$ does not depend on $r_{N+1}$, we deduce that  $h$ is unitarily equivalent to the 
Zishlin's hamiltonian, $\sum_{j=1}^N \frac{p_{j}^2}{2m}+{V}_{\rm
cl}(x)$. Therefore it is sufficient
to suppose $N\leq Z$ (cf. \cite{Zis}). Notice also that, because of the independence of ${V}_{\rm
cl}(R)$ with respect to $r_{N+1}$, one has $E_{\rm elec}=\inf\sigma(h)$.

 For a bounded below self-adjoint operator
$T$ with a ground state, $m(T)$ will denote the multiplicity of
$\inf\sigma(T)$.

Our first theorem is concerned with $H_{N,N}$

\begin{theorem}(Atoms)\ \newline\label{t4}  Assume that $N=Z$ and that the cutoff function satisfies
$(\ref{6})$ and $(\ref{40})$ then there exist $P_0>0$   and
$g_0>0$ such that, for $\va{P}\leq P_0$ and $\va{g}\leq g_0$,
$H(P)$ has a ground state such that
$m\left(H(P)\right)\leq m(h)$.
\end{theorem}
The second theorem is concerned with positive ions
$(N<Z)$

\begin{theorem}(Positive ions)\ \newline\label{t5}  Assume that $N<Z$ and that the cutoff function
satisfies $(\ref{6})$ $(\ref{40})$ and \be\label{50}\int_{\vak\leq
1}\frac{\va{\rho(k)}^2}{\vak^3} d^3 k<\infty\ee

Then there exist  $\tilde{P_0}>0$   and $\tilde{g_0}>0$ such that,
for $\va{P}\leq \tilde{P_0}$ and $\va{g}\leq \tilde{g_0}$,
$H(P)$ (with $N<Z$), has a ground state such that
$m\left(H(P)\right)\leq m(h)$.\end{theorem}

\begin{remark}\ \newline\label{r6} Theorem \ref{t4} and theorem \ref{t5} are still valid for any
operator $h$ associated with a potential $V(\rt)$   such that $h$
is essential self-adjoint on \newline$\mathcal A_N\left[
C_0^\infty(\R^{3},{\C^2})^{\otimes^N} \right]\otimes \C^d$     and
$\inf\sigma(h)$ is an isolated eigenvalue of finite
multiplicity.\end{remark}

The proof of these theorems is given in the next section. Notice
that the regularization condition $(\ref{50})$ does not allow
$\rho(k)=1$ near the origin.

A consequence of the existence of a ground state is the existence
of asymptotic Fock representations of the CCR.

For $f\in L^2(\R^3,\C^2)$, we define on $D(H_{0}(P))$ the
operators
$$
a^\sharp_{\mu,t}(f):=
e^{itH(P)}e^{-itH_{0}(P)}a^\sharp_{\mu}(f)
e^{itH_{0}(P)}e^{-itH(P)}\ .
$$
Let $Q$ be a closed null set such that the polarization vectors
$\epsilon_{\mu}(k)$ are $C^\infty$ on $\R^3\setminus Q$ for $\mu
=1,2$. We have
\begin{corollary}\ \newline\label{c7}
Suppose that the hypothesis of theorem \ref{t4} (resp. theorem
\ref{t5}) are satisfied. Then, for $f\in
C_{0}^\infty(\R^3\setminus Q)$ and for every $\Psi \in
D(H_{0}(P))$ the strong limits of $ a^\sharp_{\mu,t}(f)$
exist:
$$
\lim_{t\to \pm \infty}a^\sharp_{\mu,t}(f)\Psi
=:a^\sharp_{\mu,\pm}(f)\Psi \ .
$$
The $a^\sharp_{\mu,\pm}$'s satisfy the CCR and, if $\Phi (P)$ is a
ground state for $H(P)$, we have for $f\in
C_{0}^\infty(\R^3\setminus Q)$ and $\mu=1,2$
$$
a_{\mu,\pm}(f)\Phi(P)=0 \ .
$$
\end{corollary}

We then deduce the following corollary
\begin{corollary}\ \newline\label{c8}
    Under the hypothesis of theorem \ref{t4} (resp. theorem
\ref{t5}),
    the absolutely
    continuous  spectrum of $H(P)$  equals to $[\inf
    \sigma (H(P)), +\infty)$.
    \end{corollary}

The proofs of these two corollaries follow by mimicking
\cite{H01,H04}.

In what follows we mainly prove theorem \ref{t4}. The proof of
theorem \ref{t5} will then follow easily.

\section{Proof of theorem \ref{t4}.}\label{s4}

In this section we consider the case of atoms and thus $N=Z$.

To begin with we introduce an infrared regularized cutoff in the
interaction Hamiltonian $H_{I}(P)$. Precisely, for $\sigma
>0$, let $\rho_{\sigma}$ be a $C_{0}^\infty$ regularization of
$\rho$ such that
\begin{itemize}
    \item[(i)] $\rho_{\sigma}(k)=0$ for $\vak\leq \sigma$
    \item[(ii)] \basl{51}\lim_{\sigma \to 0}\int
    \frac{\va{\rho_{\sigma}(k)-\rho(k)}^2}{\vak^j}d^3k=0,\ j=-1,1,2.\\
    \end{split}\end{align}
\end{itemize}
We define $H_{I\sigma}(P)$ as the operator obtained from
\eqref{47} by substituting $\rho_{\sigma}(k)$ for $\rho(k)$. We
then introduce \bel{52} H_{\sigma}(P)=H_{0}(P)+H_{I\sigma}(P)
\ee and we set $E_{\sigma}(P):=\inf \sigma (H_{\sigma}(P))$.
Theorem \ref{t4} is a simple consequence of the following result
(see \cite{BFS98b})
\begin{theorem}\ \newline\label{t9}
There exist $g_{0}>0$, $\sigma_{0}>0$ and $P_0>0$ such that, for
every $g$ satisfying $\vag\leq g_{0}$, for every $\sigma$
satisfying $0<\sigma <\sigma_{0}$ and for every $P$ satisfying
$\va{P}\leq P_0$, the following properties hold:
\begin{itemize}
\item[(i)] For every $\Psi \in D(H_{0}(P))$ we have
$H_{\sigma}(P)\Psi \to_{\sigma \to 0}H(P)\Psi$ \item[(ii)]
$H_{\sigma}(P)$ has a normalized ground state $\Phi_{\sigma}(P)$
and $\esp$ is an isolated eigenvalue of finite multiplicity of
$\hsp$. \item[(iii)] Fix $\lambda \in (E_{\rm elec},\sigma_{\rm
ess}(h))$. We have \be\label{53} \langle
\Phi_{\sigma}(P),P_{(-\infty,\lambda]}\otimes P_{\Omega_{\rm ph}}
\ \Phi_{\sigma}(P)\rangle \geq 1-\delta_{g}(\lambda)\ee where
$\delta_{g}(\lambda)$ tends to zero when $g$ tends to zero and
$\delta_{g}(\lambda)<1$ for $\vag\leq g_{0}$.
\end{itemize}
\end{theorem}
Here $\sigma_{\rm ess}(h)$ is the essential spectrum of $h$.

 In
the last item, $P_{(-\infty,\lambda]}$ is the spectral projection
on $(-\infty,\lambda]$ associated to $h(b,V)$ and $P_{\Omega_{\rm
ph}}$ is the orthogonal projection on $\Omega_{\rm ph}$, the
vacuum state in $\mathcal F$.

Theorem \ref{t4} is easily deduced from theorem \ref{t9} as
follows. Let $\Phi_{\sigma}(P)$ be as in theorem \ref{t9} (ii).
Since $\norm{\Phi_{\sigma}(P)}=1$, there exits a sequence
$(\sigma_{k})_{k\geq 1}$ converging to zero and such that
$(\Phi_{\sigma_{k}}(P))_{k\geq1}$ converges weakly to a state
$\Phi(P)$. On the other hand, since
$P_{(-\infty,\lambda]}\otimes P_{\Omega_{\rm ph}}$ is finite rank
for $\lambda \in (E_{\rm elec},\sigma_{\rm ess}(h))$, it follows
from (iii) that for $\vag\leq g_{0}$ and $|P|\leq P_{0}$,
$$
\langle \Phi(P),P_{(-\infty,\lambda]}\otimes P_{\Omega_{\rm ph}}
\ \Phi(P)\rangle \geq 1-\delta_{g}(\lambda)$$ which implies
$\Phi(P)\neq 0$. Then we deduce from (i) and from a well known
result (\cite{AH97} lemma 4.9) that $\Phi_N(P)$ is a ground state
for $H(P)$.

The result concerning the multiplicity of the ground state is an
easy consequence of corollary 3.4 in \cite{H04} if $g_0$ is
sufficiently small.

So it remains to prove theorem \ref{t9}. The assertion (i) is
easily verified in section \ref{s4.1} below. The second assertion
is proved in lemma \ref{l4.8}. Actually the proof of (ii) is
lengthy but straightforward since with the infrared cutoff we have
a control of the  number of photons in term of the energy. The
real difficult part is the third one which allows to  relax the
infrared cutoff.
 The fundamental
lemma in the proof of (iii) is  lemma \ref{l11} which states that,
for $g$ and $P$ small enough, the difference
$E_{\sigma}(P-k)-E_{\sigma}(P)$ is minorized by
$-\frac{3}{4}\vak$ uniformly with respect to $\sigma$. This
estimate is essential to control the number of photons in a ground
state of $H_{\sigma}(P)$ via a pull through formula (see lemma
\ref{l12}).

\subsection{Proof of $(i)$ of theorem \ref{t9}.}\label{s4.1}

Set $\rst=\rho-\rs$.  We have

\basl{54}
&H(P)-H_{\sigma}(P)=H_{I}(P)-H_{I \sigma}(P)\\
&=-\frac{g}{M}\sum_{j=1}^N A((\A \rt)_{i},\rst)\cdot\pmd +
    \frac{Ng}{M}
A(({\A} \rt)_{N+1},\rst)\cdot\pmd\\
 &-g
\sum_{j=1}^N\frac{1}{\mu_j}\left( \sum_{k=1}^N b_{jk}A((\A
\rt)_{k},\rst)\cdot\omega_{j}\right)+Ng \sum_{j=1}^N
\frac{1}{\mu_j}b_{j,N+1}A((\A
\rt)_{N+1},\rst)\cdot\omega_{j} \\
&-\frac{g}{2m}\sum_{j=1}^N \sigma_j B((\A \rt)_{j},\rst) +
   Ng \frac{g_{\rm ncl}}{2m_{\rm ncl}}
S\cdot B(({\A} \rt)_{N+1},\rst) \\
&+\frac{1}{2M}\left(\sum_{i=1}^N g A((\A \rt)_{i},\rho)-Ng A(({\A}
\rt)_{N+1},\rho)\right)\cdot\\
&\hskip 3cm\left(\sum_{i=1}^N g A((\A \rt)_{i},\rst)-Ng A(({\A}
\rt)_{N+1},\rst)\right)\\
&+\left(\sum_{i=1}^N g A((\A \rt)_{i},\rst)-Ng A(({\A}
\rt)_{N+1},\rst)\right)\cdot\\
&\hskip 3cm\left(\sum_{i=1}^N g A((\A \rt)_{i},\rs)-Ng A(({\A}
\rt)_{N+1},\rs)\right)\\
&+ \sum_{j=1}^N\frac{1}{2\mu_j} \left( \sum_{k=1}^N b_{jk} g
A((\A\rt)_{k},\rho)- b_{j,N+1} Ng A((\A
\rt)_{N+1},\rho)\right)\cdot\\
&\hskip 3cm\left( \sum_{k=1}^N b_{jk} g A((\A\rt)_{k},\rst)-
b_{j,N+1} Ng
A((\A \rt)_{N+1},\rst)\right)\\
&+\left( \sum_{k=1}^N b_{jk} g A((\A\rt)_{k},\rst)- b_{j,N+1} Ng
A((\A \rt)_{N+1},\rst)\right)\cdot\\
&\hskip 3cm\left( \sum_{k=1}^N b_{jk} g A((\A\rt)_{k},\rs)-
b_{j,N+1} Ng A((\A \rt)_{N+1},\rs)\right).
\end{split}\end{align}

By $(\ref{51})$ we have $\lim_{\sigma \to 0}\int
    \frac{\va{\rst(k)}^2}{\vak^j}d^3k=0,$ for $j=-1,1,2$.    We then deduce from
$(\ref{34})-(\ref{39})$ , ($\ref{48})$ and from the Lebesgue's
theorem that, for every  $\psi\in D(H_{0}(P))$,
$\left(H(P)-H_{\sigma}(P)\right)\psi\rightarrow 0$ as
$\sigma$ tends to zero.

\subsection{Main estimates.}\label{s4.2} In this section we
give two lemmas concerning the function $P\mapsto\esp$.

Let $g_1>0$ such that $(\ref{40})$ is satisfied for $\vag\leq
g_1$.

\begin{lemma}\ \newline\label{l10}
 There exist  $\sigma_0\in (0,1]$ and a finite constant $C>0$ which does not
depend on $\sigma\in (0,\sigma_0]$  such that \be\label{55} E_{\rm
elec}-\vag C\leq \esp\leq E_{\rm elec} +\frac{P^2}{2M}\ee for
every $\sigma\in (0,\sigma_0]$, $P\in\R^3$ and $\vag\leq
g_1$.\end{lemma}

\proof\

Let $\varphi$ be $a$ normalized ground state of $h$ in ${\cal
A}_N\left[L^2(\R^{3},\C^2)^{\otimes^N}\right]\otimes\C^d$. Since
$\left(a_\mu(k)\om,\om\right)=
\left(\om,a_\mu(k)^\star\om\right)=0$ we have
\begin{eqnarray}\nonumber
\left\langle\hp\varphi\otimes\om,\varphi\otimes\om\right\rangle&=&
 \left\langle\hop\varphi\otimes\om,\varphi\otimes\om\right\rangle\\
 \nonumber
 &=& E_{\rm elec}+\frac{P^2}{2M}\end{eqnarray}
 and thus
\begin{eqnarray}\nonumber
\esp &:=& \inf\Bigl\{(\hsp \phi,\phi)\ \vert\ \phi\in D(\hop),\
\norm{\phi}=1\Bigr\}\\&\leq& E_{\rm
elec}+\frac{P^2}{2M}.\end{eqnarray} On the other hand, let $\ths$
be the following operator in ${\cal
A}_N\left[L^2(\R^{3},\C^2)^{\otimes^N}\right]\otimes\C^d\otimes\F:$
$$
\ths=\tho+\this
$$
with \be\label{57}\tho=h\otimes 1+1\otimes H_{\rm ph}\ee and
\basl{58} &\this=-g \skun\frac{1}{2\mu_j}\left(\sum_{k=1}^N
b_{jk}A((\A \rt)_{k},\rs)\cdot\omega_{j}+ b_{jk}\omega_{j}\cdot
A((\A
\rt)_{k},\rs)\right)\\
&+Ng \sum_{j=1}^N\frac{1}{2\mu_j}\left( b_{j,N+1}A((\A
\rt)_{N+1},\rs)\cdot\omega_{j}+ b_{j,N+1}\omega_{j}\cdot A((\A
\rt)_{N+1},\rs)\right) \\
&-\frac{g}{2m}\sum_{j=1}^N \sigma_j\cdot B((\A \rt)_{j},\rs) +
   Ng \frac{g_{\rm ncl}}{2m_{\rm ncl}}
S\cdot B(({\A} \rt)_{N+1},\rs) \\
&+ \sum_{j=1}^N\frac{1}{2\mu_j}\left( \sum_{k=1}^N b_{jk} g A((\A
\rt)_{k},\rs)- b_{j,N+1} Ng A((\A \rt)_{N+1},\rs)\right)^2
\end{split}\end{align}

One easily checks that, for $\vag\leq g_1$ , $\ths$   is a
self-adjoint operator in ${\cal
A}_N\left[L^2(\R^{3N},\C^2)^{\otimes^N}\right]\otimes\C^d\otimes\F$
with domain $D(\hop)$. Furthermore, on $D(\hop)$, we have
\basl{59} \hsp&=\ths+\frac{1}{2M}\Bigl(\pmd-g\sum_{j=1}^N A((\A
\rt)_{j},\rs)\\
&+NgA((\A \rt)_{N+1},\rs)\Bigr)^2.
\end{split}\end{align}
Hence\be\label{60}\inf\sigma(\ths)\leq\esp\ee for every
$P\in\R^3$.

By $(\ref{34})-(\ref{39})$ and $(\ref{48})$ which also hold when
$\rho$ is replaced by $\rs$ we get that there exist $\sigma_0\in
(0,1]$    and two constants $b>0$ and $a>0$ which do not depend on
$\sigma\in (0,\sigma_0]$ and $g\in [-g_1,g_1]$ and which satisfy
$bg_1<1$ such that, \be\label{61}\norm{\this\phi}\leq\vag\left(b
\norm{\tho\phi}+a\norm{\phi}\right)\ee for $\phi\in D(\tho)$ and
for $\sigma\in (0,\sigma_0]$.

 Therefore, since
$\inf\sigma(\tho)=E_{\rm elec}$, we obtain, as a consequence of
the Kato-Rellich theorem, \be\label{62}\inf\sigma(\tho)\geq E_{\rm
elec}-\max\left(\frac{a\vag}{1-b\vag},a\vag+b\vag\va{E_{\rm
elec}}\right).\ee

We then deduce the lower bound for $\esp$ with
$$
C=\max\left(\frac{a}{1-b{g_1}},a+b\va{E_{\rm elec}}\right).$$

\begin{lemma}\ \newline\label{l11} There exist $0<g_2\leq g_1$       and $P_1>0$ such that
\be\label{63} E_\sigma(P-k)-\esp\geq -\frac{3}{4}\vak\ee uniformly
for $k\in\R^3$, $\sigma\in (0,\sigma_0]$, $\vag\leq g_2$ and
$\va{P}\leq P_1$.\end{lemma}

\begin{remark}\ \newline
In this lemma we do not assume that $\esp$ is an eigenvalue of
$\hsp$ and we will use $(\ref{63})$ in Lemma \ref{l4.8} in which
we prove that $\hsp$ has a ground state.\end{remark}

\proof\ \newline

We first remark that, if $(\ref{63})$ is proved for $\hsp +C$ for
some constant $C$, it also holds for  $\hsp$  . Thus, in what
follows, we suppose $E_{\rm elec}=0$. The proof decomposes in two
steps. In the first one, we consider the large values of $\va{k}$
(namely $\vak \geq \frac{M}{7}$) while, in the second one, we
consider the small values of $\vak$ (namely $\vak \leq
\frac{M}{7}$).

From \eqref{55}, we deduce that, uniformly for
$\sigma\in(0,\sigma_0 ]$ and $\vag\leq g_{1}$, we have for all $k$
and $P$
$$
E_{\sigma}(P-k)-\esp \geq -\frac{P^2}{2M} -C\vag
$$
and thus assuming $|P|\leq {\frac{M}{\sqrt 7}}$ and $\vag \leq
\frac{M}{28C}$, \eqref{63} holds true for $\vak\geq \frac{M}{7}$.

\smallskip

Now we suppose $\vak \leq \frac{M}{7}$. As $E_{\sigma}(P-k)$
belongs to the spectrum of $H_{\sigma}(P-k)$ there exists a
sequence $(\psi_{j})_{j\geq 1}$ in $D(H_{\sigma}(P-k))$
$=D(H_{0}(0))$ such that $\norm{\psi_{j}}=1$ and
$$\lim_{j\to \infty} H_{\sigma}(P-k)\psi_{j}-E_{\sigma}(P-k)\psi_{j}
= 0\ .$$ We then have for every $j$
\begin{align}\begin{split}\label{64}
\langle H_{\sigma}(P-k)\psi_{j} , \psi_{j}\rangle &= \langle
H_{\sigma}(P)\psi_{j} , \psi_{j}\rangle +\frac{k^2}{2M}
-\frac{k}{M}\cdot\langle (P-{\rm d}\Gamma(k))\psi_{j} ,
\psi_{j}\rangle \\
&+\frac{g}{M}\sum_{i=1}^N k\cdot\langle A((\A\rt)_i,\rs)\psi_{j} ,
\psi_{j}\rangle\\
&-\frac{Ng}{M} k\cdot\langle A((\A\rt)_{N+1},\rs)\psi_{j} ,
\psi_{j}\rangle\\
&\geq \esp +\frac{k^2}{2M} - \left\vert\frac{k}{M}\cdot\langle
(P-{\rm d}\Gamma(k))\psi_{j} ,
\psi_{j}\rangle \right\vert \\
&-\frac{\vag}{M}\left\vert\sum_{i=1}^N k\cdot\langle
A((\A\rt)_i,\rs)\psi_{j} ,
\psi_{j}\rangle\right\vert\\
&-\frac{N\vag}{M} \left\vert k\cdot\langle
A((\A\rt)_{N+1},\rs)\psi_{j} , \psi_{j}\rangle\right\vert.
\end{split}\end{align}
In what follow $C$ will denote any positive constant which does
not depend on $P\in \R^3$, $k\in \R^3$, $\vag\leq g_{1}$, $\sigma
\in (0,\sigma_0 ]$ and $j\geq 1$.

 We have
\begin{align}\begin{split}\label{65}
    \left\vert\frac{k}{M}\cdot\langle (P-{\rm d}\Gamma(k))\psi_{j} ,
\psi_{j}\rangle \right|&\leq \sum_{i=1}^3
\frac{\va{k_i}}{M}\left\vert\langle
(P_i-{\rm d}\Gamma(k_i))\psi_{j} , \psi_{j}\rangle\right\vert\\
&\leq \sum_{i=1}^3
\left(\frac{{k_i}^2}{M}+\frac{{\vert{k_i\vert}}}{M}\sqrt{2M}
\norm{H_{0}(P-k)\psi_{j}}^{1/2}\right)\\
&\leq \frac{\vak^2}{M}+3\vak\sqrt{\frac{2}{M}}
\norm{H_{0}(P-k)\psi_{j}}^{1/2}\ .
\end{split}\end{align}
On the other hand, by \eqref{34}-\eqref{39} and \eqref{48}, one
shows that there exists a positive constant $C>0$ such that
\begin{align}\begin{split}\label{66}
&\frac{1}{M}\left\vert\sum_{i=1}^N k\cdot\langle
A((\A\rt)_i,\rs)\psi_{j} ,
\psi_{j}\rangle\right\vert\\
&+\frac{N}{M} \left\vert k\cdot\langle
A((\A\rt)_{N+1},\rs)\psi_{j} , \psi_{j}\rangle\right\vert\\
    &\leq C\vak(\norm{H_{\rm ph}^{1/2}\psi_{j}}+1)\\
    &\leq C \vak(\norm{H_{0}(P-k)\psi_{j}}^{1/2}+1)\ .
\end{split}\end{align}
Now, given $\epsilon>0$, let $J$ be such that \[
\norm{H_{\sigma}(P-k)\psi_{j}-E_{\sigma}(P-k)\psi_{j}}\leq
\epsilon \] for every $j\geq J$.

Inserting \eqref{65} and \eqref{66} in \eqref{64}
\begin{align}\begin{split}\label{67}
    E_{\sigma}(P-k)-\esp &\geq -\epsilon  -\frac{\vak^2}{2M}
    -3\vak\sqrt{\frac{2}{M}}\norm{H_{0}(P-k)\psi_{j}}^{1/2}\\
    &-\vak C\vag(\norm{H_{0}(P-k)\psi_{j}}^{1/2}+1).
\end{split}\end{align}
It remains to estimate $\norm{H_{0}(P-k)\psi_{j}}$.

From
$$
H_{0}(P-k)\psi_{j}=(H_{\sigma}(P-k) -E_{\sigma}(P-k)) \psi_{j}
+ E_{\sigma}(P-k)\psi_{j} -H_{I\sigma}(P-k) \psi_{j}$$ we get
for $j\geq J$ \[\norm{H_{0}(P-k)\psi_{j}}\leq \epsilon
+|E_{\sigma}(P-k)|+\norm{H_{I\sigma}(P-k) \psi_{j}}\] and we
know that there exists a positive constant $C$ such that
\be\label{69}\norm{H_{I\sigma}(P) \phi}\leq \vag
C(\norm{H_{0}(P) \phi}+1)\ee for every $P\in \R^3$, $j\geq J$ and
$g\in [-g_1,g_1]$.

Thus, choosing $\tilde{g_{1}}\leq g_{1}$ such that $\tilde{g_{1}}
C\leq \frac{1}{2}$, we get from \eqref{67} and \eqref{69} \be
\label{69b} \norm{H_{0}(P-k) \psi_{j}}\leq 2\epsilon + 2
|E_{\sigma}(P-k)| + 2\vag C \ee
 for $j\geq J$ and $\vag\leq \tilde{g_{1}}$.
 By \eqref{67} and \eqref{69b} we get
\begin{align}\begin{split}\nonumber
    E_{\sigma}(P-k)-\esp &\geq -\epsilon  -\frac{k^2}{2M}
    -6\vak\sqrt{\frac{1}{M}}(\epsilon +  |E_{\sigma}(P-k)| + \vag C)^{1/2}\\
    &-\vak C\vag((2\epsilon + 2 |E_{\sigma}(P-k)| + 2\vag C)^{1/2}+1)
\end{split}\end{align}
for every $\epsilon >0$. Hence
\begin{align}\begin{split}\label{71}
    E_{\sigma}(P-k)-\esp \geq   &-\vak\Bigl(\frac{\vak}{2M}
    +6 \sqrt{\frac{1}{M}}\left(   |E_{\sigma}(P-k)| + \vag C\right)^\usd\\
    &+ C\vag\left((2+ 2 |E_{\sigma}(P-k)| + 2\vag
    C)^\usd+1)\right)
\end{split}\end{align}
for every $P\in\R^3$ and $k\in\R^3$. From \eqref{55} we get for
$|k|\leq \frac{M}{7}$, \be\label{72}|E_{\sigma}(P-k)|\leq C\vag
+\frac{P^2}{2M} +\frac{|P|}{7}+\frac{M}{98}.\ee One then easily
shows that  there exit $P_2>0$ and $g_{2}\leq \tilde{g_{1}}$ such
that for $|P|\leq P_2$, $|k|\leq \frac{M}{7}$ and $\vag\leq
g_{2}$,
\begin{align}\begin{split}\nonumber
    E_{\sigma}(P-k)-\esp &\geq -\frac{3}{4}|k|\ .
\end{split}\end{align}

\qed

\subsection{Proof of $(iii)$ of theorem \ref{t9}}\label{s4.3}

In this section we assume that assertion $(ii)$ of theorem
\ref{t9} is already proved (see lemma \ref{l4.8}). Thus let $\psp$
denote a normalized ground state of $\hsp$, i.e.
$$
\hsp \psp = \esp \psp\ .
$$
The main problem in proving $(iii)$ of theorem \ref{t9} is to
control the number of photons in the ground state $\psp$ uniformly
with respect to $\sigma$. The operator number of photons $N_{\rm
ph}$ is given by
$$
N_{\rm ph}=\sum_{j=1,2}\int_{\R^3}d^3 k\
a^\star_{\mu}(k)a_{\mu}(k).
$$
Note that
$$
\norm{\left( 1\otimes N_{\rm
ph}^\frac{1}{2}\psp\right)}^2=\sum_{j=1,2}\int_{\R^3}d^3 k\
\norm{a_{\mu}(k)\psp}^2.
$$
We there have the following lemma
\begin{lemma}\ \newline\label{l12}
There exists a constant $C>0$ independent of $g$ and $\sigma$ such
that \be \label{73}\norm{a_{\mu}(k)\psp}\leq C\vag
\va{\rs(k)}\left(\vak^\frac{1}{2}+\frac{1}{\vak^\frac{1}{2}}
\right)\norm{(1+\va{\rt}_2)\psp} \ee for every $\sigma \in
(0,\sigma_0 ]$, $\vag \leq g_{2}$ and $|P|\leq P_2$ . Thus
\basl{74}\norm{(1\otimes N_{\rm ph}^{1/2})\psp }\leq 4C\vag
\left(\int_{\R^3}\va{\rs(k)}^2\left(
\frac{1}{\vak}+\vak\right)d^3k
\right)^{1/2}\times\\\norm{(1+\va{\rt}_2)\psp}
\end{split}\end{align} where $\va{\rt}_2$ is the Euclidian norm of
$\rt$.
\end{lemma}

\proof\

We use the gauge transformation introduced in \cite{BFS99} and
\cite{GLL}. This transformation is a particular case of the
Power-Zienau-Woolley transformation (see \cite{Coh96}, \cite{GR}).
Set
$$\tilde A(x,\rs)=A(x,\rs)-A(0,\rs)$$ and \be \label{76}
U=e^{-ig\sum_{j=1}^N r_j\cdot\left(\sum_{l=1}^N b_{jl}-N b_{j,
N+1}\right)A(0,\rs)}.\ee

We have \be\label{77} U a_\mu(k) U^\star=b_\mu(k)=a_\mu(k)+i
w_\mu(k,\rt)\ee with
$$w_\mu(k,\rt)=\frac{g}{2\pi}\frac{\rs(k)}{\vak^\usd}\emk\cdot\sjun\lef(\slun
b_{jl}-N b_{j,N+1}\ri) r_j$$

In order to estimate  $a_\mu(k)\psp$      we write
$$a_\mu(k)\psp=U^\star a_\mu(k)\pstp-i
w_\mu(k,\rt)\psp$$ where $$\pstp=U \psp.$$

In order to estimate $\norm{a_\mu(k)\pstp}$       we use the pull
through formula. We set
$$\thsp=U\hsp U^\star.$$
We have \basl{79}&\thsp =\frac{1}{2M}\lef( P-{\rm
d}\Gamma(k)-g\siun\at\lef((\A\rt)_i,\rs\ri)+gN
\at\lef((\A\rt)_{N+1},\rs\ri)\ri)^2\\
&+ \sjun\frac{1}{2\mu_j}\lef(\omega_j-\siun g
b_{ji}\at\lef((\A\rt)_i,\rs\ri)+gN b_{j,N+1}
\at\lef((\A\rt)_{N+1},\rs\ri)\ri)^2\\
&+\sum_{\mu=1,2}\int d^3k\vak b_\mu(k)^\star b_\mu(k)
+\tilde{V_{\rm el}}(\rt)\otimes 1\\
&-\frac{g}{2m}\sjun\sigma_j\cdot B\lef((\A\rt)_j,\rs\ri)+g_{\rm
ncl}\frac{Ng}{2m_{\rm ncl}} S\cdot
B\lef((\A\rt)_{N+1},\rs\ri).\end{split}\end{align}

In the first term of the r.h.s. of $(\ref{79})$ we have used
\bel{80} -g\siun A\lef(0,\rs\ri)+gN A\lef(0,\rs\ri)=0.\ee

This is only possible for atoms, i.e., when $N=Z$.

We have \bel{81}
{\tilde H_{\sigma}}(P-k)a_\mu(k)\pstp=a_\mu(k){\tilde H_{\sigma}}(P)\pstp
- \vak a_\mu(k)\pstp+ V_\mu(k)\pstp\ee

with \basl{82} &V_\mu(k)=-i
\frac{g}{2\pi}{\vak^\usd}{\rs(k)}\lef(\emk\cdot\sjun\lef(\slun
b_{jl}-N b_{j,N+1}\ri) r_j\ri)\\
&+ \frac{1}{2M}\frac{g}{2\pi}\frac{\rs(k)}{\vak^\usd}
k\cdot\emk\left(\sjun\left(\ej-\en\right)\right)\\
&+\frac{g}{\pi}\frac{\rs(k)}{\vak^\usd} \emk \cdot \frac{1}{2M}
\Biggl(P-k-{\rm d}\Gamma(k)-g\siun\at\lef(\xt_i,\rs\ri)\\
&\hskip 2cm+gN \at\lef(\xt_{N+1},\rs\ri)\Biggr)
\left(\sjun\left(\ej-\en\right)\right)\\
&+\sjun \frac{g}{\pi}\frac{\rs(k)}{\vak^\usd} \emk \cdot
\frac{1}{2\mu_j}\times\\
&\hskip 2cm\left(\omega_j-\slun g
b_{jl}\at\lef(\xt_l,\rs\ri)+gN
\at\lef(\xt_{N+1},\rs\ri)\right)\\
&\left(\slun b_{jl}\left(\el-1\right)-N
b_{j,N+1}\left(\en-1\right)\right)\\
&+ \sjun \frac{g}{2\pi}\frac{\rs(k)}{\vak^\usd} \emk \cdot
\frac{1}{2\mu_j}\times\\
&\hskip 2cm\left(\slun b_{jl}k\cdot\frac{\partial x_l}{\partial
r_j}-N b_{j,N+1}k\cdot\frac{\partial x_{N+1}}{\partial
r_j}\en\right)\\
&-\frac{g}{2m}\sjun\sigma_j\cdot\vak^\usd\rs(k)\left(\frac{k}{\vak}\wedge\emk\right)\ej\\
&\hskip 2cm+g_{\rm ncl}\frac{Ng}{m_{\rm
ncl}}S\cdot\vak^\usd\rs(k)\left(\frac{k}{\vak}\wedge\emk\right)\en,\end{split}\end{align}

where, in order to simplify the notations, we have set $$\xt_k=(\A\rt)_k=x_k -r_{N+1}$$ (the last equality is a consequence of 
\eqref{19} and shows that $\xt_k$ represents the relative position of the particule $k$ with 
respect to the center of mass).

 Thus, from $(\ref{81})$, we get \bel{83}
\left(\widetilde{H}_{\sigma}(P-k)-\esp+\vak\right)a_\mu(k)\pst=
V_\mu(k)\pst.\ee

Therefore
\bel{84}a_\mu(k)\pst=R\left(E_\sigma(P-k)-\esp+\vak\right)V_\mu(k)\pst\ee

where \bel{85}R\left(\lambda \right)=
\left(H_{\sigma}(P-k)-E_\sigma(P-k)+\lambda \right)^{-1}\ .
\ee

In order to estimate $\norm{a_\mu(k)\pst}$       we have to
estimate each term of\newline
$R\left(E_\sigma(P-k)-\esp+\vak\right)V_\mu(k)\pst$. The main
terms are those associated with the third and the fourth ones of
the r.h.s. of $(\ref{82})$

Set \basl{88}T_\mu(k)&=\\ \emk \cdot \frac{1}{2M} &\left(P-k-{\rm
d}\Gamma(k)-g\siun\at\lef(\xt_i,\rs\ri)
+gN\at\lef(\xt_{N+1},\rs\ri)\right).\end{split}\end{align}

Remark that $T_\mu(k)$ and $R$ are symmetric operator and that $ T_\mu(k)R(1)$ is bounded. 
Thus we write
\basl{89}&\Big\Vert R\left(E_\sigma(P-k)-\esp+\vak\right)
\frac{g}{\pi}\frac{\rs(k)}{\vak^\usd} T_\mu(k)\\
&\hskip 3cm\left(\sjun\left(\ej-\en\right)\right)\pst\Big\Vert\\
&\leq\sup_{\norm{\varphi}\leq
1}\frac{\vag}{\pi}\frac{\va{\rs(k)}}{\vak^\usd}\Big\vert\Bigl\langle
T_\mu(k) R\left(E_\sigma(P-k)-\esp+\vak\right)\varphi,\\
&\hskip 3cm\sjun\left(\ej-\en\right)\pst\Bigr\rangle
\Big\vert\\
&\leq\frac{\vag}{\pi}\frac{\va{\rs(k)}}{\vak^\usd}\norm{T_\mu(k)
R\left(E_\sigma(P-k)-\esp+\vak\right)}\\
&\hskip 3cm\norm{\sjun\left(\ej-\en\right)\pst}.
\end{split}\end{align}
$T_\mu(k)$    is relatively bounded with respect to
$H_{0}(P-k)-E_{\rm elec}$ with a relative bound zero. Therefore,
for $\epsilon>0$, \bel{90}
\norm{T_\mu(k)\psi}\leq\epsilon\sqrt{2M}
\norm{\left(H_{0}(P-k)-E_{\rm
elec}\right)\psi}+C_\epsilon\norm{\psi}\ee

for some finite constant $C_\epsilon>0$   and for $\psi\in
D\left(H_{0}(0)\right)$.

In order to estimate $\left(H_{0}(P-k)-E_{\rm elec}\right)\psi$, we write
 \beal{91}\left(H_{0}(P-k)-E_{\rm elec}\right)\psi&=&
\left(H_{\sigma}(P-k)-\esp+\vak\right)\psi\\
&+&\left(\esp-\vak\right)\psi-H_{I\sigma}\psi.\eea

Thus

\basl{92}\norm{\left(H_{0}(P-k)-E_{\rm elec}\right)\psi}&\leq
\norm{\left(H_{\sigma}(P-k)-\esp+\vak\right)\psi}\\
&+\left(\vert\esp\vert+\vak\right)\norm{\psi}+\norm{H_{I\sigma(P-k)}\psi}.\end{split}\end{align}

Note that, by $(\ref{55})$, there exists a finite constant $M>0$
such that  \newline$\vert\esp\vert\leq M$     for
$\sigma\in(0,\sigma_0]$.

By the theorem \ref{t2} there exist two finite constants
$\alpha>0$ and $\beta>0$ such that $\alpha g_1<1$          and
\bel{93}\norm{H_{I\sigma(P-k)}\psi}\leq \vag\left(\alpha
\norm{\Big(H_{0}(P-k)-E_{\rm
elec}\right)\psi}+\beta\norm{\psi}\Big)\ee for $\vag\leq g_1$,
$\sigma\in(0,\sigma_0]$ and
 every $P$ and $k$.

 We get, for some finite constants $C>0$,
\basl{94}\norm{\left(H_{0}(P-k)-E_{\rm elec}\right)\psi}&\leq
C\norm{\left(H_{\sigma}(P-k)-\esp+\vak\right)\psi}\\
&+\left(\vak+1\right)\norm{\psi}\end{split}\end{align}

and by $(\ref{90})$,

\basl{95}\norm{T_\mu(k)\psi}&\leq
C\norm{\left(H_{\sigma}(P-k)-\esp+\vak\right)\psi}\\
&+\left(\vak+1\right)\norm{\psi}.\end{split}\end{align}

Now in view of \eqref{89}, we would like to apply \eqref{95} with \newline
  $\psi=R\left(E_\sigma(P-k)-\esp+\vak\right)\varphi$. First we remark that,
according to Lemma \ref{l11}, we have
\bel{86}E_\sigma(P-k)-\esp+\vak\geq\frac{\vak}{4}\ee for
$\sigma\in(0,\sigma_0]$, $\vag\leq g_2$, $\va{P}\leq P_2$ and thus we
get
\bel{87}\norm{R\left(E_\sigma(P-k)-\esp+\vak\right)}\leq\frac{4}{\vak}\ .\ee

Therefore combining   $(\ref{87})$ and $(\ref{95})$ we obtain

\bel{96}\norm{T_\mu(k)R\left(E_\sigma(P-k)-\esp+\vak\right)\varphi}\leq
C\left(1+\frac{1}{\vak}\right)\norm{\varphi}\ee

for every $\varphi\in {\cal
A}_N\left[L^2(\R^{3},\C^2)^{\otimes^N}\right]\otimes\C^d\otimes\F$.

Thus

\bel{97}\norm{T_\mu(k)R\left(E_\sigma(P-k)-\esp+\vak\right)}\leq
C\left(1+\frac{1}{\vak}\right).\ee

We easily show that \basl{98}
\norm{\sjun\left(\ej-\en\right)\pstp}&\leq C\vak \norm{(1+\vert \xt
\vert_2)\pstp}\\ &\leq C\vak \norm{(1+\vert \rt
\vert_2)\pstp}.\end{split}\end{align}

We then get for the third term of \eqref{82},

\basl{99} &\Big\Vert R\left(E_\sigma(P-k)-\esp+\vak\right)
\frac{g}{\pi}\frac{\rs(k)}{\vak^\usd} \emk
\cdot \frac{1}{2M} \Big(P-k-{\rm
d}\Gamma(k)\\&\hskip 2cm -g\siun\at\lef(\xt_i,\rs\ri)
+gN\at\lef(\xt_{N+1},\rs\ri)\Big)\\&\hskip 2cm
\left(\sjun\left(\ej-\en\right)\right)\pst\Big\Vert\\
&\leq C\frac{\vag}{\pi}\vert\rs(k)\vert
\left(\vak^\usd+\frac{1}{\vak^\usd}\right)\norm{(1+\vert \rt
\vert_2)\pstp}\end{split}\end{align}

for  $\sigma\in(0,\sigma_0]$, $\vag\leq g_2$, $\va{P}\leq P_2$.

Similarly, we have for the fourth term of \eqref{82}

\basl{100} &\Big\Vert R\left(E_\sigma(P-k)-\esp+\vak\right)
\sjun\frac{g}{\pi}\frac{\rs(k)}{\vak^\usd} \emk \cdot
\frac{1}{2\mu_j}\\&\hskip 2cm\times \left(\omega_j-\slun g
b_{jl}\at\lef(\xt_l,\rs\ri)+gN
\at\lef(\xt_{N+1},\rs\ri)\right)\\
&\hskip 1cm\times\left(\slun b_{jl}\left(\el-1\right)-N
b_{j,N+1}\left(\en-1\right)\right)\pstp\Big\Vert\\
&\hskip 4cm\leq C\frac{\vag}{\pi}\vert\rs(k)\vert
\left(\vak^\usd+\frac{1}{\vak^\usd}\right)\norm{(1+\vert \rt
\vert_2)\pstp}.\end{split}\end{align}

It is easy to verify that the remaining terms of the r.h.s. of,
$(\ref{84})$ associated to the remaining ones of $(\ref{82})$ are
also bounded by

$$
C\frac{\vag}{\pi}\vert\rs(k)\vert
\left(\vak^\usd+\frac{1}{\vak^\usd}\right)\norm{(1+\vert \rt
\vert_2)\pstp}.
$$

This concludes the proof of the lemma \ref{l12}.

\qed

Let us remark that the above proof is a little bit formal because
of the use of the «  Pull Through » formula. But, by mimicking
\cite{H04} one easily gets a rigorous proof. We omit the details.

The following Lemma allows us to control $\norm{(1+\vert \rt
\vert_2)\pstp}$. Let $\delta={\rm dist}(E_{\rm
elec},\sigma(h)\backslash\E_{\rm elec}>0\}$. By $(\ref{55})$,
there exists $P_3>0$ and $0<g_3\leq g_2$ such that \basl{101}
\esp&\leq E_{\rm elec}+\frac{\delta}{3},\ {\rm for\ }\va{P}\leq
P_3 {\rm\ and\ for\ }\sigma\in (0,\sigma_0]\\
C\vag&\leq \frac{\delta}{12},\ {\rm for\ }\vag\leq
g_3\end{split}\end{align}

where $C$ is the constant in $(\ref{55})$.

Let $\Delta$ be an interval such that $\esp\in \Delta$ for
$\va{P}\leq P_3$ and for $\sigma\in (0,\sigma_0]$ and $\sup \Delta
< E_{\rm elec}+\frac{\delta}{2}$.

Thus \bel{102} E_{\rm elec}+\frac{2\delta}{3}-\sup \Delta -C\vag
\geq \frac{\delta}{12} \ee for $|P|\leq P_3$ and $\vag\leq g_{3}$

Let $\eta>0$ be such that \bel{103} 0<\eta^2<E_{\rm elec}
+\frac{2\delta}{3}-\sup \Delta -C\vag\ . \ee

We then have
\begin{lemma}\label{l14}\ \newline
 There exists a finite constant $M_{\Delta}>0$ such that
  \be \label{104} \norm{(e^{\eta
|\rt|_2}\otimes 1)\psp}\leq M_{\Delta}\ee for $|P|\leq P_3$,
$\vag\leq g_{3}$ and $\sigma\in (0,\sigma_0]$.
\end{lemma}

The proof of lemma \ref{l14} easily follows by mimiking the proof
of theorem II.1 in \cite{BFS98b}.

We denote by $P_{(.]}$ the spectral measure of $h$ and by
$P_{\Omega_{\rm ph}}$ the orthogonal projection on $\Omega_{\rm
ph}$. We have the following lemma
\begin{lemma}\ \newline\label{l4.6}
Fix $\lambda \in (E_{\rm elec},\inf\sigma_{\rm ess}(h))$. There
exists $\delta_{g}(\lambda)>0$ such that $\delta_{g}(\lambda)\to
0$ when $g\to 0$ and \be \label{105} \langle P_{[\lambda
,\infty)}\otimes P_{\om} \psp \ ,\ \psp \rangle \leq
\delta_{g}(\lambda) \ee for every $\sigma \in (0,\sigma_0 ]$,
$|P|\leq P_3$ and $\vag \leq g_{3}$.
\end{lemma}
\proof\    Since $P_{\om}H_{\rm ph} =0$ and $P_{\om}(P-{\rm
d}\Gamma(k))^2= P^2 P_{\om}$ we get
\begin{align}\begin{split}\label{106}
    (&P_{[\lambda ,\infty)}\otimes P_{\om})(\hsp -\esp)= P_{[\lambda
    ,\infty)}(h\otimes I)\otimes P_{\om} \\
    &+ (\frac{P^2}{2M}-\esp)P_{[\lambda ,\infty)}\otimes P_{\om}+
    P_{[\lambda ,\infty)}\otimes P_{\om}\hisp\ .
\end{split}\end{align}
Applying this last equality to $\psp$ we get
\begin{align}\begin{split}\label{107}
    0&=P_{[\lambda
    ,\infty)}(h\otimes I)\otimes P_{\om}\psp \\
    &+ (\frac{P^2}{2M}-\esp)P_{[\lambda ,\infty)}\otimes P_{\om}\psp\\
    &+
    P_{[\lambda ,\infty)}\otimes P_{\om}\hisp\psp\ .
\end{split}\end{align}
Since $hP_{[\lambda ,\infty)}\geq \lambda P_{[\lambda ,\infty)}$
we obtain from lemma \ref{l11} and \eqref{54}

 \basl{108}
    \langle P_{[\lambda ,\infty)}&\otimes P_{\om} \psp \ ,\ \psp
    \rangle \leq \\
    &\frac{1}{E_{\rm elec}-\lambda}\langle (P_{[\lambda ,\infty)}
    \otimes P_{\om})\hisp \psp \ ,\ \psp
\rangle
\end{split}\end{align}
for every $\sigma \in (0,\sigma_0 ]$, $|P|\leq P_3$ and $\vag \leq
g_{3}$. The lemma then follows from \eqref{108} and \eqref{93}.

\qed

\medskip

We are now able to conclude the proof of $(iii)$ of theorem
\ref{t9}. We have \basl{109}
   \langle P_{(-\infty,\lambda ]}\otimes P_{\om} \psp \ &,\ \psp\rangle
    =\\
    &1-\langle P_{[\lambda ,\infty)}\otimes P_{\om} \psp \ ,\ \psp
    \rangle \\
    &-\langle 1\otimes P^\perp_{\om} \psp \ ,\
\psp
    \rangle \  \ .
\end{split}\end{align}

The second term in the r.h.s. of $(\ref{109})$ is estimated by
lemma \ref{l4.6} and, noticing that $P^\perp_{\om}\leq N_{\rm
ph}$, the two last terms are estimated by lemma \ref{l12} and
lemma \ref{l14}. Theorem \ref{t4} then follows by choosing
$P_0=\inf(P_1,P_2,P_3)$ and $g_0=\inf(g_1,g_2,g_3)$ and from the
following Lemma

\qed

\begin{lemma}\ \newline\label{l4.8}
$H_{\sigma}(P)$ has a ground state for $0<\sigma \leq\sigma_0$,
$|P|\leq P_0$ and $\vag \leq g_{0}$.
\end{lemma}

In this lemma we prove the assertion (ii) of theorem \ref{t9} :
for $\sigma$ and $P$ small enough, the Hamiltonian with infrared
cutoff has a ground state. This result is not surprising but the
complete proof is long. Actually it follows by mimicking
\cite{FGSray,FGScom,DG99} (see also \cite{AGG} and \cite{Mo})
 and, here, we only sketch the proof.

First we are faced with the lack of smoothness of the
$\epsilon_{\mu}(k)$'s which define  vector fields on spheres
$|k|=$cst (see \cite{LL,G04}). It suffices to consider one
example. From now on suppose that
$$
\epsilon_{1}(k)=
\frac{1}{\sqrt{k_{1}^2+k_{2}^2}}(k_{2},-k_{1},0)\quad \mathrm{ and
}\quad \epsilon_{2}(k)= \frac{k}{\vak}\wedge \epsilon_{1}(k)\ .
$$
The functions $\epsilon_{\mu}(k)$, $\mu =1,2$, are smooth only on
$\R^3\setminus \{(0,0,k_{3})\mid k_{3}\in \R \}$. Nevertheless, in
our case, we can overcome this problem easily by choosing the
regularization $\rho_{\sigma}$ of $\rho$ as a $C^\infty$ function
whose support does not intersect the line $\{(0,0,k_{3})\mid
k_{3}\in \R \}$. From now on we suppose that it is the case.

Let $\omod$ be the modified dispersion relation as defined in
(\cite{FGScom}, section 5, hypothesis 3), i.e. : $\omod$ is a
smooth function satisfying
\begin{itemize}
  \item[(i)] $\omod \geq \max (\vak, \frac{\sigma}{2})$ for all $k\in
  \R^3$, $\omod =\vak $ for $\vak \geq \sigma$.
  \item[(ii)] $ |\nabla \omod| \leq 1$ for all $k\in
  \R^3$, and $\nabla \omod \neq 0$ unless $k=0$.
  \item[(iii)] $\omega_{\mathrm{mod}}(k_{1}+k_{2})\leq
  \omega_{\mathrm{mod}}(k_{1})+ \omega_{\mathrm{mod}}(k_{2})$ for all
  $k_{1},k_{2} \in \R^{3}$.
    \end{itemize}
    We set
$$
H_{ph,\mathrm{mod}}= \sum_{\mu =1,2}\int \omod
a^\star_{\mu}(k)a_{\mu}(k)d^3k
$$
and let $H_{\rm mod,\sigma}(P)$ be the same Hamiltonian as in
(\ref{45}) except that in (\ref{46}) we replace $H_{\rm ph}$ by
$H_{ph,mod}$.

Theorem \ref{t2}, with the same assumption \eqref{40}, is still
valid for $\hmsp$. Set $\emsp := \inf \sigma (\hmsp)$. Then
$\emsp$ still satisfies lemma \ref{l11} and \eqref{63} for the
same constants $g_{2}$ and $P_1$. Moreover, according to
(\cite{FGScom}; thm 3), $\esp =\emsp$ for $|P|\leq P_1$ and
$\vag\leq g_{2}$ and $\esp$ is an eigenvalue of $\hsp$ if and only
if $\emsp$ is an eigenvalue of $\hmsp$. Thus in order to prove
that $\hsp$ has a ground state it suffices to prove that $\emsp <
\inf \sigma_{\mathrm{ess}}(\hmsp)$. The proof is by contradiction.

It follows by proving that, if $\emsp=\inf\sigma_{\rm
ess}(\hmsp)$, then we can construct a sequence of states
$(\phi_n)$ whose energy converges to $\emsp$   and that have a non
vanishing component along the delocalized photons and therefore
has an energy which is larger than  $\emsp +\frac{\sigma}{2}$
(c.f. \cite{FGScom} or \cite{AGG} for details).

\section{Proof of theorem \ref{t5}.}\label{s5}

 When $N<Z$ the above proof only fails in
lemma \ref{112} where we used $N=Z$ in order to obtain \eqref{80} and thus \eqref{79}.

 We now haveThis implies that, now, instead of \eqref{85}, we have to consider
 \basl{110}T_\mu(k)=&\emk \cdot \frac{1}{2M}\times\\
&\left(P-k-{\rm d}\Gamma(k)-g\siun A\lef((\A\rt)_i,\rs\ri) +gZ
A\lef((\A\rt)_{N+1},\rs\ri)\right).\end{split}\end{align}

We cannot cancel the singularity  $\frac{1}{\vak^\usd}$ in
$A(\cdot,\rs)$ as we have done when $N=Z$ by substituting
$\at(\cdot,\rs)$ for $A(\cdot,\rs)$ and by applying the unitary
transformation $U$ now given by

 \be \label{111}
U=e^{-ig\sum_{j=1}^N r_j\cdot\left(\sum_{k=1}^N b_{jk}-Z b_{j,
N+1}\right)A(0,\rs)}.\ee

Therefore, in order to estimate all the terms associated with
$T_\mu(k)$, we have to suppose that \bel{112} \int_{\vak\leq
1}\frac{\va{\rs(k)}^2}{\vak^3} d^3 k<\infty\ee for
$\sigma\in(0,\sigma_0]$.

Thus we get \bel{113}\norm{(1\otimes N_{\rm ph}^{1/2})\psp }\leq
C\vag \left(\int_{\R^3}\va{\rs(k)}^2\left(
\frac{1}{\vak^3}+\vak\right)d^3k \right)^{1/2}\norm{\psp} \ee for
$\sigma \in (0,\sigma_0 ]$,  $\vag \leq g_{2}$ and $|P|\leq P_2$.
We then conclude the proof of theorem \ref{t5} as above by
choosing by choosing $P_0=\inf(P_1,P_2)$ and $g_0=\inf(g_1,g_2)$.

\qed

\section{The hydrogen atom in a constant magnetic field.}\label{s6}

 We consider a hydrogen atom in $\R^3$ interacting
    with a classical magnetic field $B_0$ pointing along the $x_{3}$-axis and
    with a quantized electromagnetic field.

The index $1$ is related with the electron while the index $2$ is
related with the proton. So we introduce in the Hilbert space
$$
\mathcal H_{{\rm magn}}=L^2(\R^3)\otimes L^2(\R^3)\otimes\F.
$$
The hamiltonian is the following one:
\begin{align}\begin{split}\nonumber
H_{{\rm magn}}&=\frac{1}{2m_1}\left( p_1-\frac{e}{2} B_0\wedge
x_1-g
A(x_1,\rho)\right)^2\\
&+\frac{1}{2m_2}\left( p_2+\frac{e}{2} B_0\wedge x_2+g
A(x_2,\rho)\right)^2\\
&+V(x_1-x_2)\otimes 1 + 1\otimes H_{\rm ph}.
\end{split}\end{align}

Here for simplicity, we omit spins and we still replace the charge
$e$ in front of $A(\cdot,\rho)$ by the parameter $g$.  $V$ denote
the coulomb potential: $V(x)=-e^2\backslash \vert{x}$.

The total momentum of the system is given by $P=K\otimes
1+1\otimes {\rm d}\Gamma(k)$ where
$$K=p_1+\frac{e}{2} B_0\wedge x_1+p_2-\frac{e}{2} B_0\wedge x_2.
$$

One easily verifies that the $3$ components of $P$ commute with
$H_{{\rm magn}}$ and that $[P_j,P_k]=0$   for $j,k=1,2,3$.
Therefore $H_{{\rm magn}}$ admits a decomposition as a direct
integral over the joint spectrum of $(P_1,P_2,P_3)$.

\begin{remark} \ \newline In fact  $K\wedge K=-i B_0 (e_1+e_2)$  and
 the three components of $P$
commutes together only when the total charge is zero. This explain
the difference with the case of the electron that we considered in
\cite{AGG} for which we were able to reduce the hamiltonian only
with respect to the third component of the total momentum.
\end{remark}

In order to obtain a simpler reduced hamiltonian we prefer first
to transform $H_{{\rm magn}}$    in such a way the momentum $K$ is
transformed into $P_R$, the conjugate momentum of the center of
mass $R$.

Considering the unitary transformation
$$
U=e^{i\frac{e}{2} r\cdot B_0\wedge R}
$$
with $r=x_1-x_2$, one obtains
$$U^{-1} K U=P_R$$
and denoting  $\widetilde{H}_{{\rm magn}}=U^{-1}H_{{\rm magn}}U$
one gets
\begin{align}\begin{split}\nonumber
\widetilde{H}_{{\rm magn}}&=\frac{1}{2m_1}\left(
\frac{m_1}{M}P_R+p_r-\frac{e}{2} B_0\wedge r-g
A(x_1,\rho)\right)^2\\
&+\frac{1}{2m_2}\left( \frac{m_2}{M}P_R-p_r-\frac{e}{2} B_0\wedge
r+g
A(x_2,\rho)\right)^2\\
&+V(r)\otimes 1 + 1\otimes H_{\rm ph}
\end{split}\end{align}
where $p_r=-i\nabla_r$.

We have
$$[P_R,\widetilde{H}_{{\rm magn}}]=0.$$

Therefore the hamiltonian $\widetilde{H}_{{\rm magn}}$  ,
unitarily equivalent to ${H}_{{\rm magn}}$, admits a decomposition
as a direct integral over the joint spectrum of the three
components of $\widetilde{P}=P_R\otimes 1+1\otimes {\rm
d}\Gamma(k)$. We obtain formally
\[\widetilde{H}_{{\rm magn}}\ \simeq\ \int_{\R^3}^\oplus 
\widetilde{H}_{{\rm magn}}(P)\,\,d^3 P
\]
on
\[
\mathcal H_{{\rm magn}}\ \simeq\ \int_{\R^3}^\oplus
L^2(\R^{3})\otimes\F\,\, d^3P
\]

with

\begin{align}\begin{split}\nonumber
\widetilde{H}_{{\rm magn}}(P)&=\frac{1}{2m_1}\left(
\frac{m_1}{M}\pmd+p-\frac{e}{2} B_0\wedge r-e
A\left(\frac{m_2}{M}r,\rho\right)\right)^2\\
&+\frac{1}{2m_2}\left( \frac{m_2}{M}\pmd -p-\frac{e}{2} B_0\wedge
r+e
A\left(-\frac{m_1}{M}r,\rho\right)\right)^2\\
&-\frac{e^2}{r}\otimes 1 + 1\otimes H_{\rm ph},
\end{split}\end{align}
where $p=p_r$.

Precisely, as usual we replace the charge $e$ by a parameter $g$ in 
the interaction terms and write
$$
\widetilde{H}_{{\rm magn}}(P)=\widetilde{H}_{{\rm
Hyd},0}(P)+\widetilde{H}_{{\rm magn},I}(P)
$$
with, introducing the electronic hamiltonian
$$h(B_0)=\frac{1}{2m_1}\left(p-\frac{e}{2} B_0\wedge
r\right)^2+\frac{1}{2m_2}\left( p+\frac{e}{2} B_0\wedge
r\right)^2-\frac{e^2}{r}\ ,
$$
one has
$$
{H}_{{\rm magn},0}(P) = h(B_{0})\otimes 1+ \frac{1}{2M}\pmd^{2} + 1\otimes H_{\rm ph}.
$$
and
\begin{align}\begin{split}\nonumber
\widetilde{H}_{{\rm magn},I}(P)&=\frac{g}{M}
\pmd \left( A\left(-\frac{m_1}{M}r,\rho\right)-A\left(\frac{m_2}{M}r,\rho\right)\right)\\
&+\frac{g^2}{2m_1} A\left(\frac{m_2}{M}r,\rho\right)^2+
\frac{g^2}{2m_2} A\left(-\frac{m_1}{M}r,\rho\right)^2\\
&-\frac{g}{M} \pmd B_0\wedge r \ .
\end{split}\end{align}

From \cite{AHS81} we know that the electronic hamiltonian
$h(B_0)$ has an isolated ground state
$\varphi_{\rm el}$. Hence, for $|P|< M$, $\varphi_{\rm el}\otimes\om$
is the ground state of $\widetilde{H}_{{\rm
magn},0}(P)$ and remains isolated. Then following the same steps as in
section \ref{s4} with straightforward adaptation of lemmas
\ref{l10}, \ref{l11} and \ref{l12} one obtains the following
theorem

\begin{theorem}\label{t16}\ \newline Assume that the cutoff function 
satisfies
$(\ref{6})$. Then there exist $P_0$  and $g_0>0$ such
that for $\va{P}\leq P_0$ and $\va{g}\leq
g_0$, $\widetilde{H}_{{\rm magn}}(P)$  has a ground state with
$$m\left(\widetilde{H}_{{\rm magn}}(P)\right)
\leq m\left(h(B_0))\right).$$
\end{theorem}

\qed

\bibliography{qed}
\bibliographystyle{abbrv}
\end{document}